\documentclass{aa}
\usepackage{graphicx}
\usepackage{lscape}

\begin{document}

\def\kms{km~s$^{-1}$}
\def\cm2{cm$^{-2}$}
\def\hkpc{$~h_{70}^{-1}$ kpc}
\def\icm{cm$^{-2}$}
\def\lya{Ly$\alpha$}
\def\lyb{Ly$\beta$}
\def\lyc{Ly$\gamma$}
\def\mgtwo{Mg~II}
\def\sifour{Si~IV}
\def\cfour{C~IV}
\def\apm{APM08279+5255}

\title{The Sizes and Kinematic Structure of
Absorption Systems Towards the Lensed Quasar \apm\thanks{Based on 
{\it Hubble Space Telescope} data from program 8626}.}

\titlerunning{Sizes and Structure of QSO Absorption Systems.}

   \author{Sara L. Ellison\inst{1,2,3}
          \and
          Rodrigo Ibata\inst{4}
          \and
          Max Pettini\inst{5}
          \and
           Geraint F. Lewis\inst{6}
          \and
         Bastien Aracil\inst{7}
          \and
          Patrick Petitjean\inst{7}
          \and
          R. Srianand\inst{8}
          }

   \offprints{S. Ellison}

  \institute{
        P. Universidad Cat\'olica de Chile, Casilla 306, Santiago 22, Chile\\
        \and
	European Southern Observatory, Casilla 19001, Santiago 19, 
        Chile\\
	\and
	Current address: Department of Physics and Astronomy,
	University of Victoria,  Elliott Building, 3800 Finnerty Rd.,
	Victoria, BC, V8P 1A1, Canada.\\
        \email{sarae@uvic.ca}\\
        \and
        Observatoire de Strasbourg, 11, rue de l'Universit\'e, F-67000, 
        Strasbourg, France\\
        \and
        Institute of Astronomy, Madingley Rd, Cambridge, CB3 0HA, U.K.\\
        \and
         Institute of Astronomy, School of Physics, A~29, University of Sydney, NSW 2006, Australia\\
        \and
        Institut d'Astrophysique de Paris -- CNRS, 98bis Boulevard 
        Arago, F-75014 Paris, France\\
        \and
        IUCAA, Post Bag 4, Ganeshkhind, Pune 411 007, India}

   \date{Received / Accepted}

\abstract{We have obtained spatially resolved spectra of the 
$z_{\rm em}=3.911$ triply imaged
QSO \apm\ using the Space Telescope Imaging  Spectrograph (STIS)
on board the {\it Hubble Space Telescope} ({\it HST}).  
We study the line of sight equivalent width (EW) 
differences and velocity shear of high and low ionization 
absorbers (including a damped Lyman alpha [DLA] system identified
in a spatially unresolved ground based spectrum) in the three 
lines of sight.  The combination of
a particularly rich spectrum and three sight-lines 
allow us to study 27 intervening absorption systems
over a redshift range $1.1 < z_{\rm abs} < 3.8$, 
probing proper transverse dimensions of 30 $h_{70}^{-1}$\,pc up to 
2.7\,\hkpc.  
We find that high ionization systems (primarily C~IV absorbers) do not 
exhibit strong 
EW variations on scales $<0.4$\, \hkpc; their fractional EW differences 
are typically less than 30\%.  
When combined with previous work on other QSO pairs,
we find that the fractional variation increases steadily
with separation out to at least $\sim 100$\,\hkpc.  Conversely, low
ionization systems (primarily Mg~II absorbers) show strong
variations (often $> 80$\%) over kpc scales.  A minimum radius for 
strong (EW\,$> 0.3$\,\AA) Mg~II systems of $> 1.4$\,\hkpc\
is inferred from absorption coincidences in all lines of sight.
For weak Mg~II absorbers (EW\,$< 0.3$\,\AA), 
a maximum likelihood analysis indicates a most probable coherence 
scale of 
2.0\,\hkpc\ for a uniform spherical geometry, with 95\% confidence limits
ranging between 1.5 and 4.4\,\hkpc.  
The weak \mgtwo\ absorbers may therefore represent a distinct population
of smaller galaxies compared with the strong \mgtwo\ systems which 
we know to be associated with 
luminous galaxies whose halos extend over tens of kpc.  Alternatively,
the weak systems may reside in the outer parts of larger galaxies,
where their filling factor may be lower.
By cross-correlating spectra along different lines of sight, we infer shear
velocities of typically less than 20\,\kms\ for both high and low
ionization absorbers.  Finally, for systems with
weak absorption that can be confidently converted to column densities,
we find constant $N$(C~IV)/$N$(Si~IV) across the three lines of sight.  
Similarly, the [Al/Fe] ratios in the $z_{\rm abs} = 2.974$ DLA 
are consistent with solar relative abundances 
over a transverse distance of $\sim 0.35$\,\hkpc.
\keywords{quasars: absorption lines -- 
quasars: individual: APM08279+5255 -- galaxies: ISM -- galaxies: 
structure -- galaxies: kinematics and dynamics -- gravitational lensing}
}

\maketitle

\section{Introduction}

The study of high redshift absorption line systems in the lines of
sight (LOS) to lensed and multiple QSOs can yield information on
the sizes of intervening galaxies and the structure of the 
intergalactic medium (IGM).
Statistical 
analyses of the \lya\ forest in multiple LOS
have established that the coherence length of \lya\ absorbers 
ranges from a few pc to hundreds of kpc
(e.g. Crotts 1989; Smette et al 1995; Petry, Impey \& Foltz 1998;
Petitjean et al. 1998; Lopez, Hagen \& Reimers 2000; Rauch et al. 2001b).
A significant fraction of \lya\ forest clouds with neutral
hydrogen column densities greater than $\log N$(H~I)$ > 14$
(where $N$(H~I) is measured in cm$^{-2}$)
have associated metals detected primarily via
highly ionized species such as C~IV,
even out to very high redshifts (Ellison et al 2000; Songaila 2001;
Pettini et al. 2003).  From studies of small
scale variations using lensed QSOs, these high ionization
metal systems have been found to exhibit little structure
on transverse scales of a few pc (Rauch, Sargent \& Barlow 1999).
However, on kpc scales variations can be seen between in some absorption 
components (e.g. Smette et al 1995; Rauch, Sargent \& Barlow 2001a);  recently
Tzanavaris \& Carswell (2003) have 
inferred sizes of $\sim 0.5 - 8$\,\hkpc\ for individual C~IV clouds
\footnote{As usual $h_{70}$ is the Hubble constant in units
of 70\,km~s$^{-1}$~Mpc$^{-1}$}.
Observations of multiple QSOs extend this type of
analysis to larger scales and indicate that coherence (i.e. coincidence)
between C~IV systems still exists over
dimensions of $\sim$ 100\,\hkpc\ (e.g. Petitjean et al 1998;
Lopez, Hagen \& Reimers 2000).   
A complementary study of C~IV absorbers
in the vicinity of Lyman break galaxies has shown that the 
regions where metals are dispersed into the IGM  
probably extend over hundreds of kpc
(Adelberger et al. 2003).

Compared with the results of the
\lya\ forest and high ionization absorbers, the structural properties
of the galactic halos probed by low ionization metal line systems
are relatively few.  This is mainly due to the
fact that low ionization absorbers have a lower incidence 
per unit redshift than \lya\ or C~IV systems.  
Rauch et al. (2002) probed four pairs of low ionization systems
and found significant spatial and column density differences for
individual components on scales of a few hundred pc, although in every 
case the absorber covered both LOS indicating that low ionization
halos have overall dimensions larger than $\sim 0.5$\,kpc.  
Kobayashi et al. (2002) studied the probable damped \lya\ system
(DLA) at $z_{\rm abs} = 2.974$ towards 
\apm\ and again found structural differences in the Mg~II lines 
over scales of $\sim 0.3$\,kpc\footnote{The HI profile of the DLA has 
only been studied in a ground-based, spatially unresolved spectrum (Petitjean
et al. 2000).  In these data, the N(HI) is poorly constrained 
and the actual column density may fall slightly below the canonical
limit of 2$\times10^{20}~\rm{cm}^{-2}$, although we will continue to 
refer to it herein as a DLA.  Since the STIS data do not cover the \lya\
of this absorber, we have no information on the N(HI) for the separate
LOS.}.  Using a spatially unresolved
HIRES spectrum of \apm\ Petitjean et al. (2000) inferred
sub-kpc sizes for some \mgtwo\ systems based on partial coverage arguments.
Churchill et al (2003) found evidence for minor structural 
differences on even smaller scales, although these are not
reflected in discernable chemical abundance variations
over distances of $0.135$\,kpc in a  DLA
at $z_{\rm abs} = 1.3911$.  Indirect determinations of the size
of \mgtwo\ systems, based on impact parameters and number density per 
unit redshift, indicate dimensions $\sim$ 50 \hkpc\ (Steidel 1995;
Bergeron and Boiss\'e 1991).
Such large sizes have been claimed even for relatively weak absorbers
(Churchill \& LeBrun 1998; Churchill et al 1999).

Taken together, these observations indicate that 
the coherence length of C~IV systems is much larger
than that of Mg~II systems. This is qualitatively 
consistent with the simple picture---drawn more than
a decade ago from the incidence of these absorbers per unit 
redshift---of clumpy, low ionization, gas embedded in
larger, more homogeneous, and more highly ionized
outer halos (e.g. Steidel 1993).
However, the number of multiple QSOs in which
the transverse spatial dimensions of 
metal lines (particularly low ionization systems)
have been determined is still relatively small.  
Moreover, the range of linear scales probed is somewhat limited, 
largely because ground-based spectroscopy requires image separations 
typically greater than 0.8\, arcsec.

\begin{figure}
\centerline{\rotatebox{180}{\resizebox{6.5cm}{!}
{\includegraphics{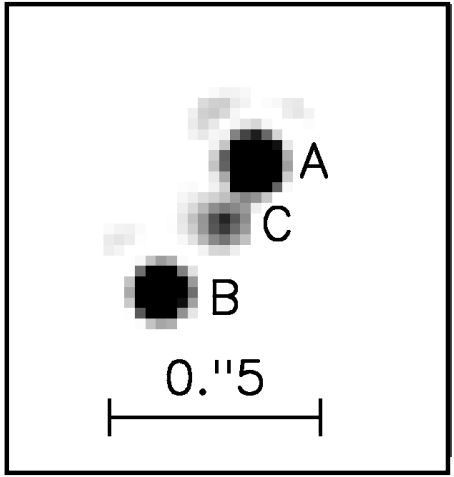}}}}
\caption{\label{nic}NICMOS image of \apm\
showing the three lensed images of the QSO
(adapted from Ibata et al. 1999).
The F110W magnitudes of the three components are
as follows: A = 13.45$\pm0.02$, B = 13.74$\pm0.02$, C = 15.37$\pm0.03$\,.
}
\end{figure}

Here we present new spatially resolved observations of the 
triply imaged QSO $z_{\rm em} = 3.911$ \apm\ obtained with 
the Space Telescope Imaging Spectrograph (STIS)
on board the {\it Hubble Space Telescope} ({\it HST}).
The unique triple nature of this system has been recently
confirmed by Lewis et al. (2002b) using a subset of these STIS
data and can be explained as lensing by a naked cusp, probably
associated with a highly inclined disk (Lewis et al. 2002a).
In Figure \ref{nic} we reproduce a NICMOS image of \apm\ obtained
by Ibata et al. (1999), illustrating the configuration of the
lensed images. \apm\ is a particularly powerful case for
probing the transverse dimensions of QSO absorbers for
two reasons. First, its triple nature provides us with 
three times as many baselines as the more common QSO pairs.
Second, its absorption spectrum is particularly rich
in intervening (as well as associated) systems
(Ellison et al 1999a,b), including a DLA at 
$z_{\rm abs} = 2.974$ (Petitjean et al. 2000).
The number of cases in which DLAs have been probed 
with multiple LOS is still relatively small (see Smette et al. 1995;
Zuo et al 1997; Boiss\'e et al. 1998; Lopez et al. 1999; Gregg et al. 2000;
Churchill et al. 2003; Wucknitz et al. 2003).

In this paper we adopt a $\Omega_M = 0.3$, $\Omega_{\Lambda}=0.7$, 
$H_0 = 70$\,km~s$^{-1}$~Mpc$^{-1}$ cosmology throughout, unless
otherwise stated. In calculating the transverse distances 
corresponding to the angular separations of the lensed images
of \apm, we adopt a redshift $z_{\rm lens} = 1.062$ for the lensing
galaxy which has so far remained undetected. This is the redshift 
of a strong, low ionization, absorption system revealed by the
HIRES spectrum of the QSO (Petitjean et al. 2000).

\section{Observations}

Observations with STIS on {\it HST} were obtained over the 
period November 2001--November 2002.
Over a total of 25 orbits, we used five different settings of the 
G750M grating to obtain complete coverage 
over the wavelength region $\sim 6000-8600$\,\AA.  
With the 0.2\,arcsec wide slit the dispersion
is 0.55\,\AA~pixel$^{-1}$ and the spectral resolution
1.6\,\AA\ FWHM (80 -- 55\,\kms).
Table \ref{obs_table} summarises the observations.
The acquisition was achieved by aligning the slit along the bright 
components A and B and then off-setting perpendicularly by 
0.02\, arcsec to centre on the faint, spatially
intermediate image C.  
Between each exposure the targets were stepped
along the slit by 1\,arcsec.  

In this paper, we also make use of the ground-based echelle spectrum of
\apm\ obtained with HIRES on the Keck~I telescope (obtained in April and
May 1998).  A detailed
description of those observations (and associated data reduction)
can be found in Ellison et al. (1999a).  In brief, we obtained a total
of 8.75 hours of spectroscopy for \apm\ with complete wavelength
coverage between 4400 and 9250\,\AA.  The spectral resolution of the data is
6\,\kms\ and the typical signal-to-noise ratio is S/N\,$\simeq 100$ 
per pixel.  The three QSO images are spatially
{\it un}resolved in these spectra.
The data are publicly available via an ftp server at the 
Institute of Astronomy, Cambridge 
(ftp://ftp.ast.cam.ac.uk/pub/papers/APM08279).

\section{Data Reduction}

\begin{figure*}
\centerline{\rotatebox{270}{\resizebox{11cm}{18cm}
{\includegraphics{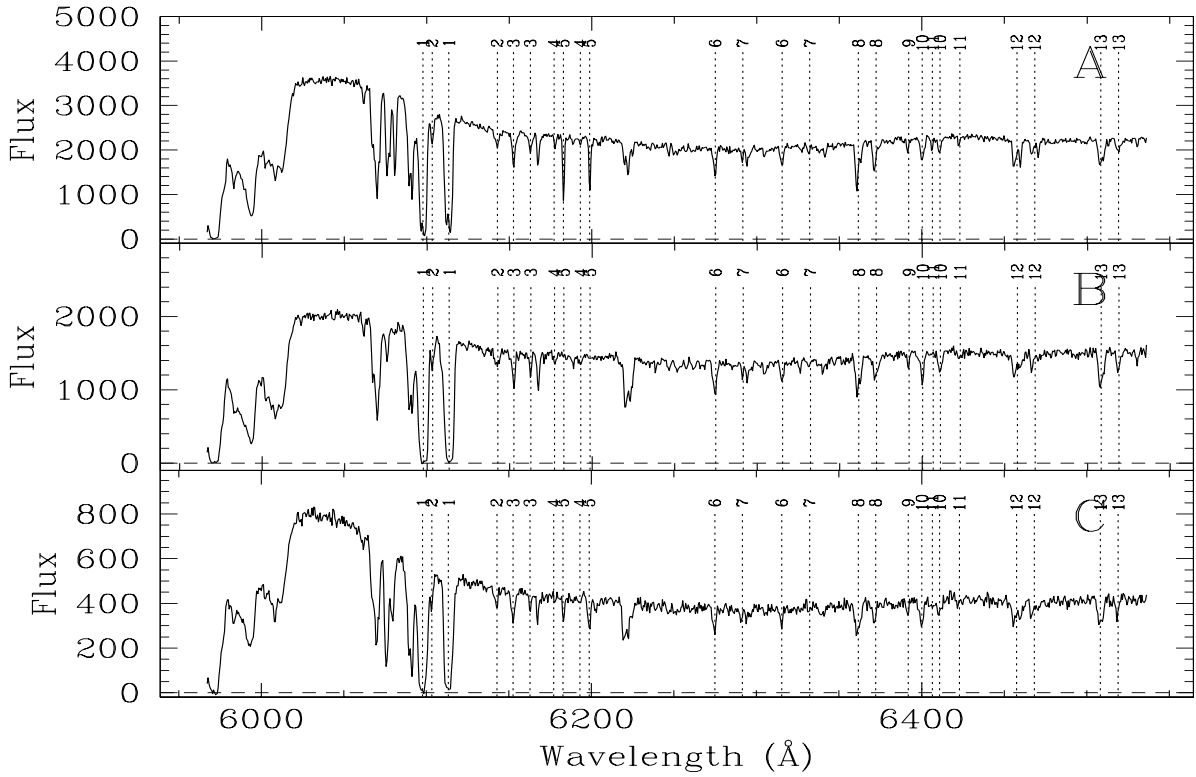}}}}
\caption{STIS spectra (arbitrary flux units) of the 3 LOS with the main 
absorption systems marked.
Tick marks refer to systems listed in Table \ref{system_table}.}
\end{figure*}

\begin{figure*}
\addtocounter{figure}{-1}
\centerline{\rotatebox{270}{\resizebox{11cm}{18cm}
{\includegraphics{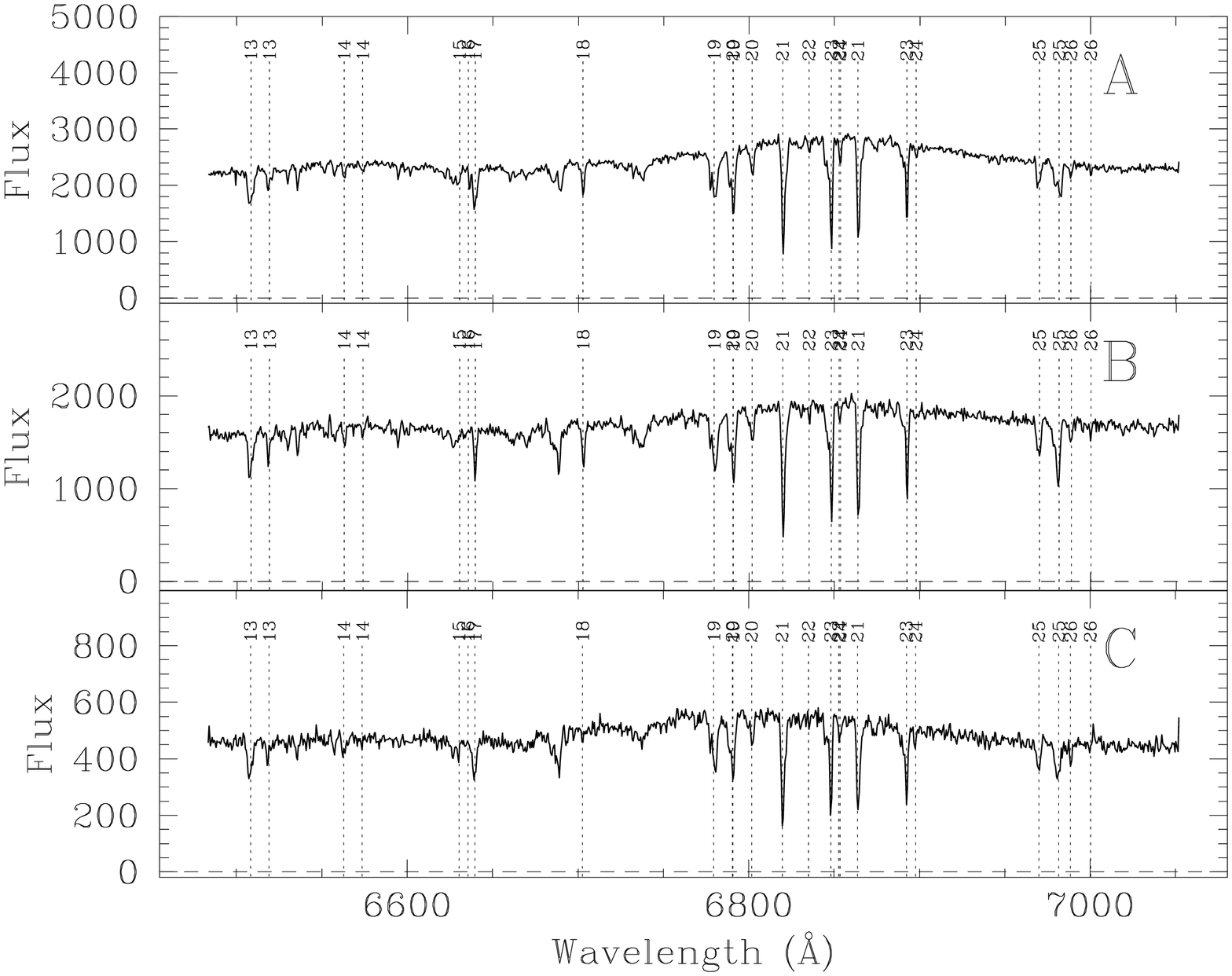}}}}
\caption{\label{stis_spec}Contd. }
\end{figure*}

\begin{figure*}
\addtocounter{figure}{-1}
\centerline{\rotatebox{270}{\resizebox{11cm}{18cm}
{\includegraphics{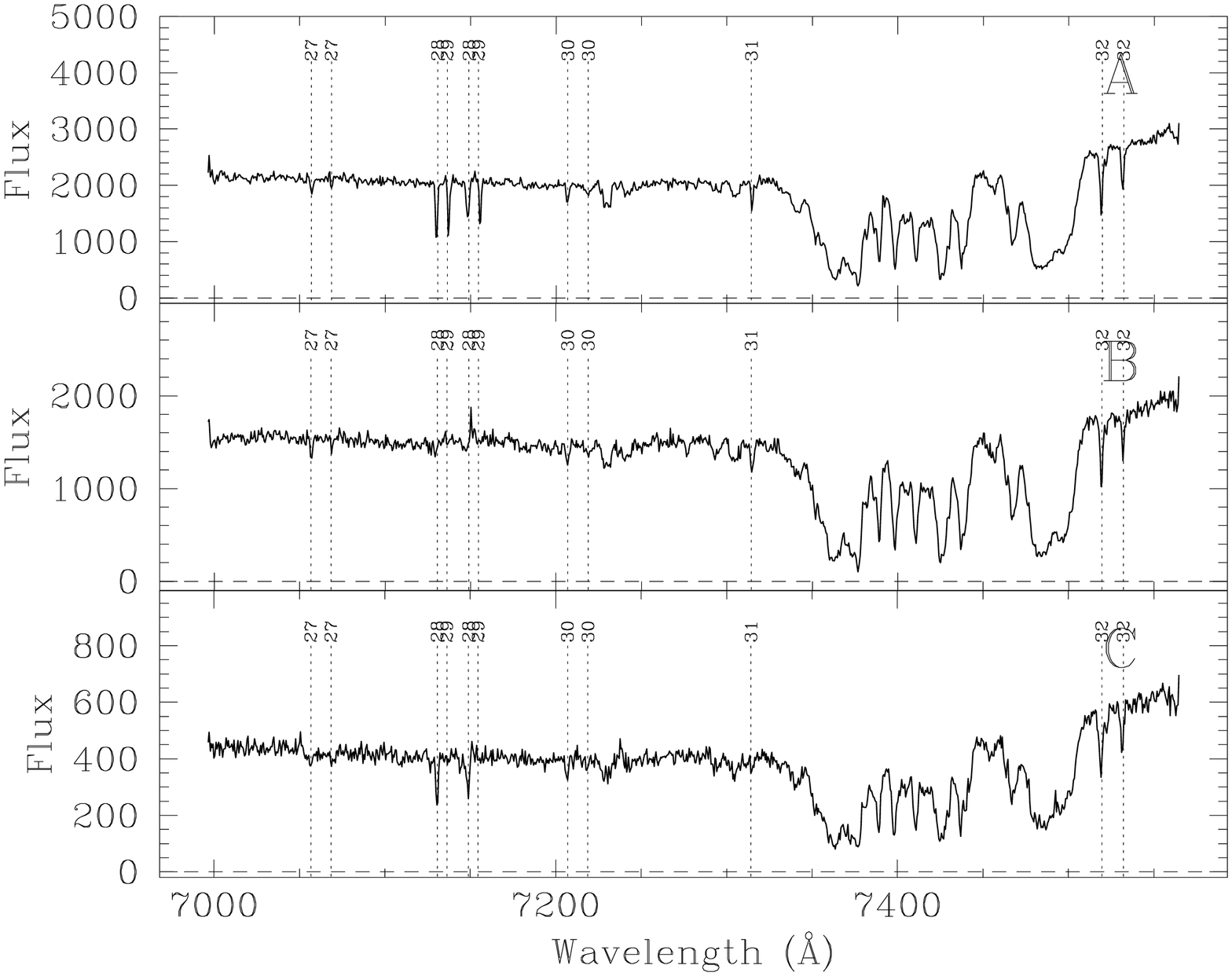}}}}
\caption{Contd.}
\end{figure*}

\begin{figure*}
\addtocounter{figure}{-1}
\centerline{\rotatebox{270}{\resizebox{11cm}{18cm}
{\includegraphics{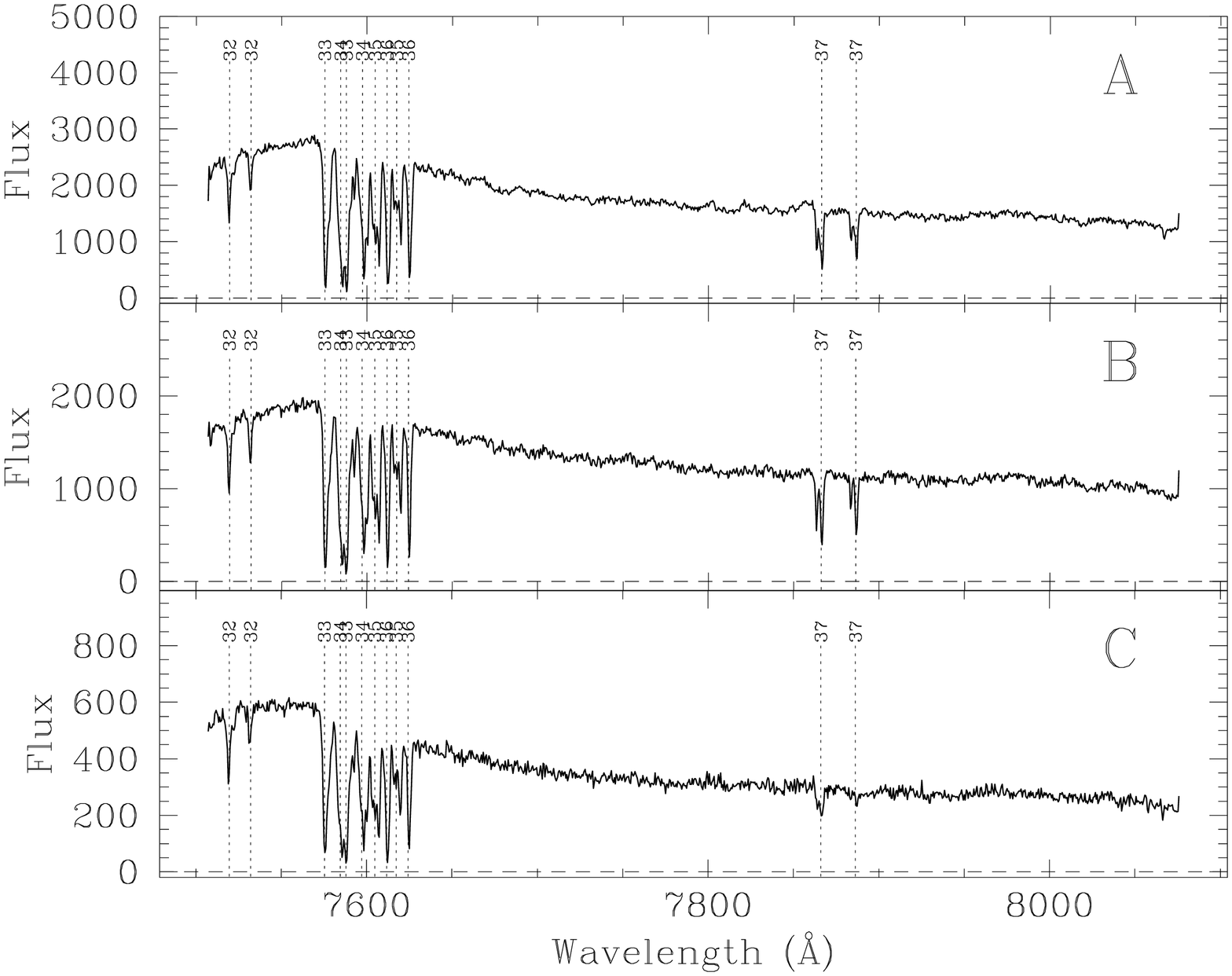}}}}
\caption{Contd.}
\end{figure*}

\begin{figure*}
\addtocounter{figure}{-1}
\centerline{\rotatebox{270}{\resizebox{11cm}{18cm}
{\includegraphics{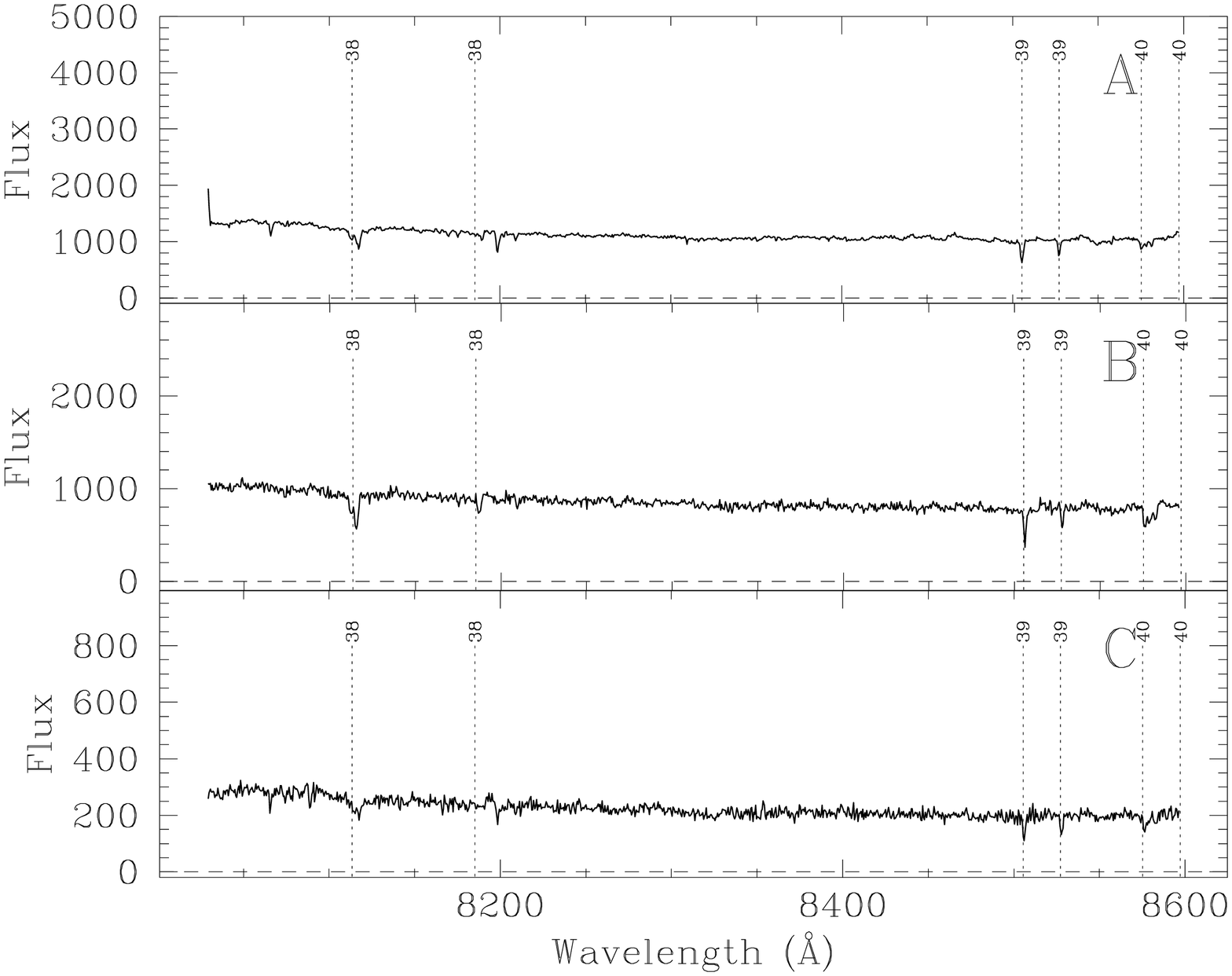}}}}
\caption{Contd.}
\end{figure*}

The STIS data were reduced as described by Lewis et al. (2002b).
Despite the fact that the HIRES and STIS spectra were all mapped to
a common vacuum heliocentric wavelength scale,
we found that there remained some systematic
velocity differences, internally between each of the three STIS spectra 
(for the three lines of sight), and also between the STIS and HIRES
spectra. We suspect that this is caused by the 
slight offset from the centre of the slit of images A and B.
The small shifts ($<20$\,\kms) were corrected by 
performing a cross correlation between each STIS spectrum 
and the HIRES spectrum and applying the corresponding
velocity adjustment.  Note that this procedure, which 
corrects for the overall \textit{systematic} offsets
in the wavelength scales, should not affect the relative
velocity shear of individual absorption systems.

Each STIS spectrum was normalised by fitting a cubic spline through
a smoothed spectrum and using the HIRES spectrum as a guide to the
continuum regions.  This was particularly important in STIS settings
where absorption lines are abundant, for example the $\lambda 6252$ setting.
Equivalent widths for each absorption system were measured and 
errors estimated for counting statistics and continuum placement.

The final (un-normalised) extracted spectra are shown in 
Figure \ref{stis_spec}.  The differences in broad emission line
strengths have been previously pointed out by Lewis et al. (2002b) and
are probably due to microlensing effects.  Tick marks indicate
the main absorption systems, and Table \ref{system_table} 
identifies the species and
redshifts.  A total of 40 transitions identified from the HIRES spectrum
were detected in the STIS data, including 10 Mg~II systems (three of which
are also detected in Fe~II) and 16 C~IV systems
(five of which also exhibit Si~IV absorption).  
We also detect Fe~II and Al~II 
in the DLA at $z_{\rm abs}=2.974$  and
Ca~II at $z_{\rm abs}=1.062$ in what is likely to be the lensing galaxy
(Petitjean et al. 2000).  There remain several unidentified lines
in the HIRES spectrum (also present in the STIS data).

\section{Equivalent Width Analysis}

Due to the comparatively low resolution of the STIS spectra
(FWHM\,$\simeq 55-80$\,\kms), 
it is not possible to determine column densities directly 
by Voigt profile fitting of the absorption lines; thus
we generally work here with the line equivalent widths.

We first examine the consistency between the STIS 
and HIRES measurements of equivalent width (EW).
The fractional contribution to the total flux for the three components has
been measured from NICMOS images to be 0.51$\pm0.02$, 0.40$\pm0.02$ and 
0.09$\pm0.03$ for A, B 
and C respectively (Ibata et al. 1999---see Figure 1).  
The corresponding fractional contributions 
to the EWs in the spatially unresolved HIRES spectrum 
can therefore be written as

\begin{equation}\label{ew_eqn}
EW_{HIRES}= 0.51 EW_A + 0.40 EW_B + 0.09 EW_C
\end{equation}

\begin{figure}
\centerline{\rotatebox{270}{\resizebox{6.5cm}{!}
{\includegraphics{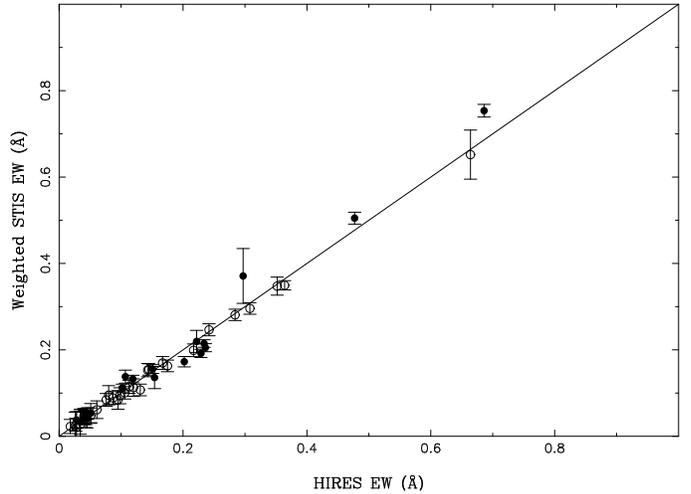}}}}
\caption{\label{ew_check}Comparison of HIRES EWs and STIS EWs weighted by 
their respective flux contributions as given in equation \ref{ew_eqn}.
In all the figures in this paper, solid points
indicate low ionization systems, such as Mg~II
and Fe~II, as well as the DLA
at $z_{\rm abs} = 2.974$.  Open points denote high 
ionization systems such as C~IV and Si~IV.  Error bars are 1$\sigma$.
}
\end{figure}

Figure \ref{ew_check} shows the consistency check 
using equation (\ref{ew_eqn}). 
All of the lines are found to be in
very good agreement, with the exception of two 
Mg~II systems.  System number 40 apparently shows some
discrepancy between the ground-based and HST data.  However, 
the HIRES spectrum is strongly affected by sky lines in this
wavelength region and the system lies at the very limit of the
STIS spectral coverage.
The EW determinations consequently have unusually large errors 
(see Figure 2 and Table 2) and are actually only deviant at about 
the 1$\sigma$ level.  The other anomaly is the case of 
Mg~II $\lambda 2796$ at $z_{\rm abs}=1.813$ (number 37) which 
has similar values of EW in A and B, but is much weaker in C.  
In principle, such EW inconsistencies between 
the spatially resolved STIS spectra
and the unresolved HIRES data may arise from 
a combination of QSO variability (Lewis, Robb \& Ibata 1999) 
and time delays between the different LOS.  
However, there is no sign of EW variations between different 
epoch STIS data (taken a year apart), which is probably not
surprising given the microlensing timescale is $\sim$ 40 years. 
Furthermore, other absorbers 
that do not fully cover all three sources do not
show such deviations from the HIRES values.  
Since the other member of the Mg~II doublet, $\lambda 2803$,
and two Fe~II transitions in the same absorption system
(Fe~II~$\lambda 2600$ and $\lambda 1608$)
all agree with the HIRES data, it is most likely that 
we have simply underestimated the measurement error 
for the stronger Mg~II$\lambda$2796 line.
Thus, overall, there is excellent agreement between the 
new data presented here and the earlier measurements
by Ellison et al. (1999b).

\begin{figure}
\centerline{\rotatebox{270}{\resizebox{6.5cm}{!}
{\includegraphics{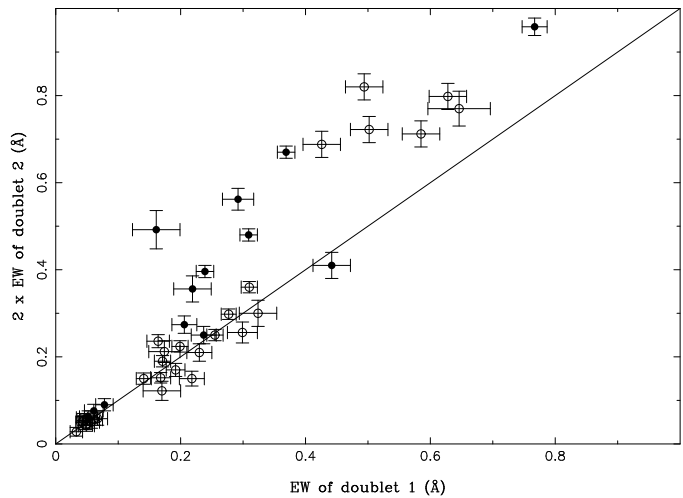}}}}
\caption{\label{doublets}Test for line saturation by comparing the
rest frame EWs of doublet pairs of C~IV and \mgtwo\ systems from the STIS 
spectra. 
On the $x$-axis we have plotted the value of equivalent
width for the stronger component of each doublet, while
on the $y$-axis we have plotted the EW of the weaker
component multiplied by 2 to account for the lower value
of oscillator strength.
Deviation from a one-to-one relation above EW $\sim$ 0.2 \AA\ 
indicates the onset of saturation for lines stronger than this
limit.
}
\end{figure}

A straightforward conversion between equivalent width and column 
density is possible for unsaturated absorption systems located on
the linear part of the curve of growth (i.e. for lines which are 
optically thin).  
Although we have the HIRES
spectrum as a guide to which systems may be saturated, partial coverage 
effects may make strong lines appear unsaturated
if they are only present in one or two of the LOS (or if they vary
significantly between the LOS).   
In Figure \ref{doublets} we plot the rest frame EWs of
doublet components in the STIS data, multiplying the
weaker component by 2 (to account for the
lower value of the oscillator strength).
For lines above $\sim$ 0.2 \AA, we note the onset of a deviation from a 
one-to-one relation in some systems, indicative of possible saturation.  
Therefore, we adopt the conservative approach of only estimating
column densities for absorbers with EW $<$ 0.2 \AA.  This method
is vindicated by the very good agreement ($<<0.1$ dex) between 
the flux-weighted column densities from the STIS data (analogous to 
Figure \ref{ew_check} and Eqn. \ref{ew_eqn})
and the Voigt profile fitted column densities of
the HIRES data (e.g. Ellison et al. 2000).

\section{Equivalent Width and Column Density Variations Between Sightlines}

We now take advantage of the high spatial
resolution of the STIS/{\it HST} combination
to examine equivalent width differences between
the sight-lines to the three images of \apm.
We distinguish between high (e.g. C~IV, Si~IV)
and low (e.g. Mg~II, Fe~II) ionization systems, since there is evidence
that the characteristic scales of these absorbers are different
(e.g. Rauch et al. 1999, 2002).  
Thus, in the plots that follow 
we use throughout solid symbols for low ionization
systems and open symbols for high ionization systems.
It is important to realise that, since the rest-frame wavelengths
of the transitions of the high ions are generally lower than those of 
the low ions, the two sets of absorption lines probe quite different
redshifts (and therefore transverse spatial scales) in the data 
analysed here.
We also note that there are a number of high ionization transitions 
at redshifts $z_{\rm abs} \sim z_{\rm em}$ 
(e.g. absorber numbers 21, 23, 24 and 32--36 
in Tables \ref{system_table} and \ref{ew_table}).
In the following analysis we have not included absorption systems within
5000~\kms\ of the QSO emission redshift because their locations along
the lines of sight cannot be deduced if their redshifts are not cosmological.
For absolute variations, we consider only systems with 
EW\,$ < 0.2$\,\AA\ in the stronger member of the doublet,
so that we can convert to column density 
($N$, measured in atoms cm$^{-2}$) assuming that
the lines are on the linear part of the curve of growth, where

\begin{equation}
N = 1.13 \times 10^{20} \frac{EW (\rm\AA)}{\lambda_0^2 f}
\end{equation}

where $f$ is the oscillator strength and $\lambda_0$ is the rest
wavelength of the transition in \AA.

\subsection{High Ionization Systems}

\begin{figure}
\centerline{\rotatebox{270}{\resizebox{6.5cm}{!}
{\includegraphics{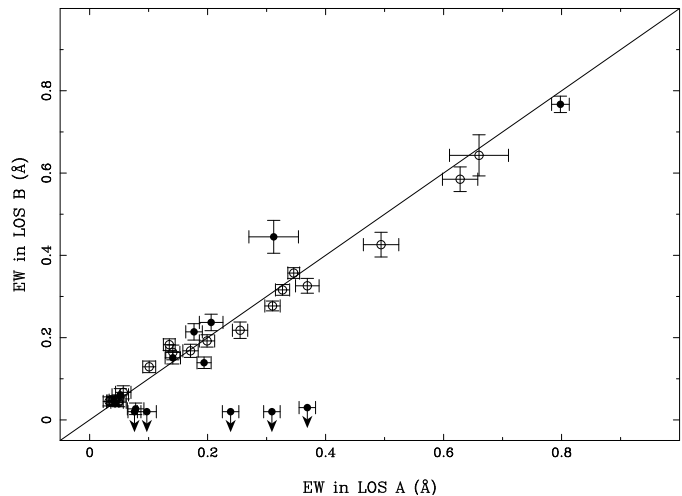}}}}
\caption{\label{EW_AB}
Comparison of rest frame EWs between lines of sight A and B.  Limits
in this and all subsequent figures are 3$\sigma$.
}
\end{figure}

\begin{figure}
\centerline{\rotatebox{270}{\resizebox{6.5cm}{!}
{\includegraphics{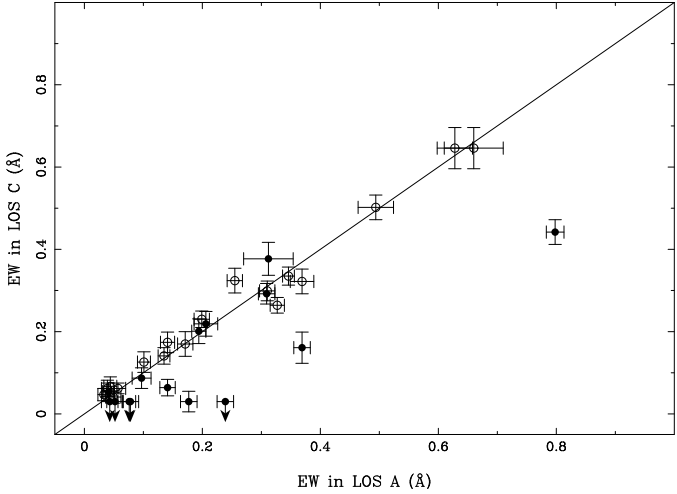}}}}
\caption{\label{EW_AC}As for Figure \ref{EW_AB} but for LOS A and C.}
\end{figure}

We identify a total of 16 high ionization systems
with LOS separations  0.028--0.340\,\hkpc.   
This is an excellent complement at small scales to the
study of high ionization systems by Rauch et al. (2001a)
who probed scales up to 5\,kpc but with only a few 
systems at separations 
less than 0.1\,kpc.\,\footnote{Rauch et al. adopted a
$\Omega_M = 1$, $\Omega_{\Lambda} = 0$,
$H_0 = 50$\,km~s$^{-1}$~Mpc$^{-1}$ cosmology,
while we use today's `consensus' values of these
parameters, $\Omega_M = 0.3$, $\Omega_{\Lambda} = 0.7$,
$H_0 = 70$\,km~s$^{-1}$~Mpc$^{-1}$.
However, the LOS separations as a function of redshift
differ by less than 10\% between the two cosmologies 
(at the redshifts considered here) and we have therefore not
applied any correction to the values reported by Rauch et al. (2001a)
quoted in the text.}

\begin{figure}
\centerline{\rotatebox{270}{\resizebox{6.5cm}{!}
{\includegraphics{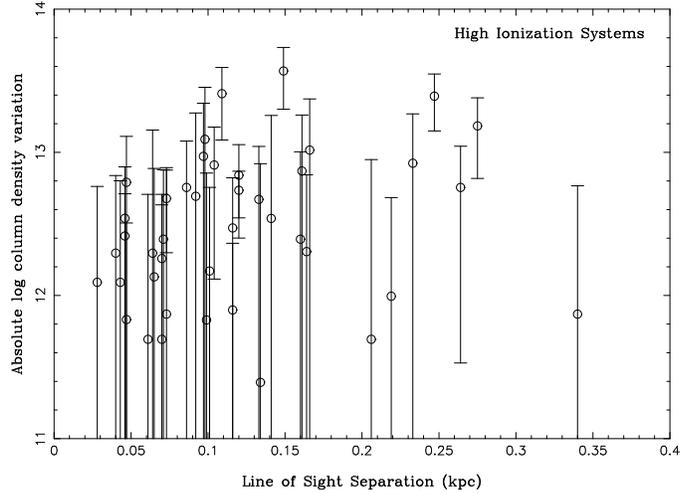}}}}
\caption{\label{abs_high}Absolute column density differences for high 
ionization systems as a function of proper LOS separation}
\end{figure}

\begin{figure}
\centerline{\rotatebox{270}{\resizebox{6.5cm}{!}
{\includegraphics{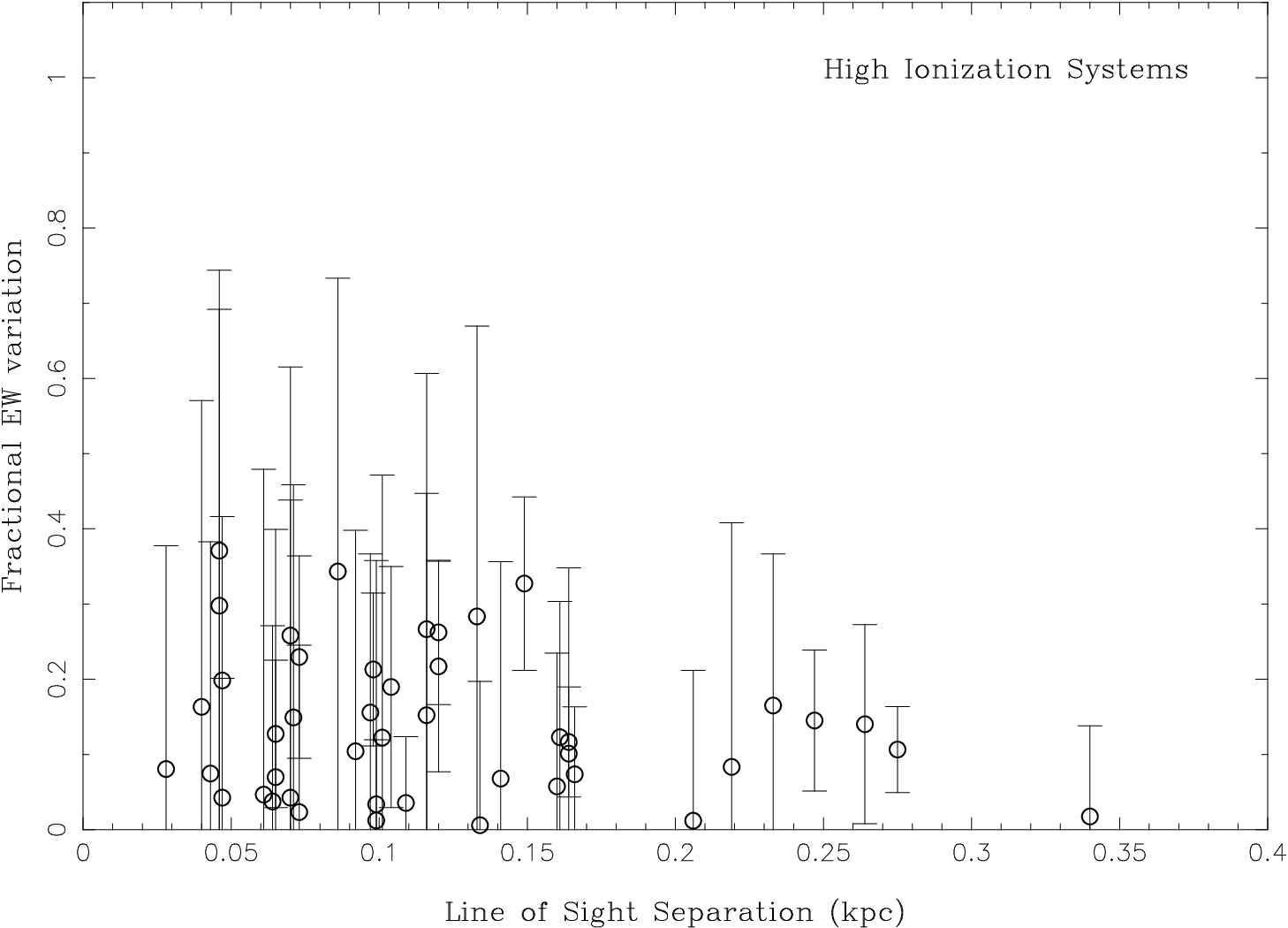}}}}
\caption{\label{frac_high}Fractional EW differences for high ionization
systems as a function of proper LOS separation}
\end{figure}

In  Figures \ref{EW_AB} and
\ref{EW_AC} we compare the equivalent widths between LOS A--B (largest
separation) and A--C (smallest separation) respectively. 
In all cases the equivalent widths of high ionization systems
show only mild differences between the different LOS, 
indicating that the structures which give rise to them 
are relatively uniform over scales of a few hundred pc.
C~IV equivalent width differences between sight-lines generally amount to 
less than 50\,m\AA, which corresponds to column densities
$\log N$(C~IV)$ < 13.1$, see Figure \ref{abs_high}.
Figure \ref{frac_high} shows the fractional EW variation 
between sightlines as a function of LOS separation 
(all three A--B, B--C and A--C pairs are included).
Typically the fractional equivalent width difference 
$\Delta EW=\frac{\mid~EW_1-EW_2~\mid}{max(EW_1,EW_2)}$ 
is less than 30\%, although the errors can be large. 
There is no clear trend between variation and LOS 
separation on all scales below $\sim 0.4$\,kpc.  
Presumably the suggestion of a possible decrease in 
$\Delta$~EW for LOS $<0.1$\,kpc in the data by 
Rauch et al. (2001a) was an artifact of small number statistics,
since we see no evidence of it in the more extensive 
set of measurements presented here 
(although see below and Figure \ref{frac_all}).

So far we have considered only whole absorption systems,
without reference to their multi-component structure.
While we do see hints of variations on a component-by-component
basis even with our (limited) spectral resolution 
(one example is system 12---see Figure \ref{stis_spec}),
we will investigate such differences in more detail in a future
paper by reconstructing the high resolution line profiles
(Aracil et al., in preparation).

\subsection{Low Ionization Systems}

We identify a total of 17 low ionization systems (mostly Mg~II and Fe~II)
with LOS separations 0.41--3.09\,\hkpc.  This list includes 
Fe~II~$\lambda 1608$ and Al~II~$\lambda 1670$
associated with the DLA at $z_{\rm abs} = 2.974$
previously studied by Petitjean et al. (2000).  
As can be appreciated from inspection of Figures \ref{EW_AB} and 
\ref{EW_AC}, there can be significant LOS differences
in the absorption lines from low ionization gas, 
with several instances where absorption
is detected in only one or two of the three sight-lines 
\footnote{Four out of five of the non-detections in LOS B compared with LOS A 
involve one absorption system at $z_{\rm abs} = 1.55$
(line numbers 15, 16, 28 and 29 in Table \ref{ew_table}).
}.

\begin{figure}
\centerline{\rotatebox{270}{\resizebox{6.5cm}{!}
{\includegraphics{0003.f9.ps}}}}
\caption{\label{abs_low}Absolute column density differences for low ionization
systems as a function of proper LOS separation}
\end{figure}

\begin{figure}
\centerline{\rotatebox{270}{\resizebox{6.5cm}{!}
{\includegraphics{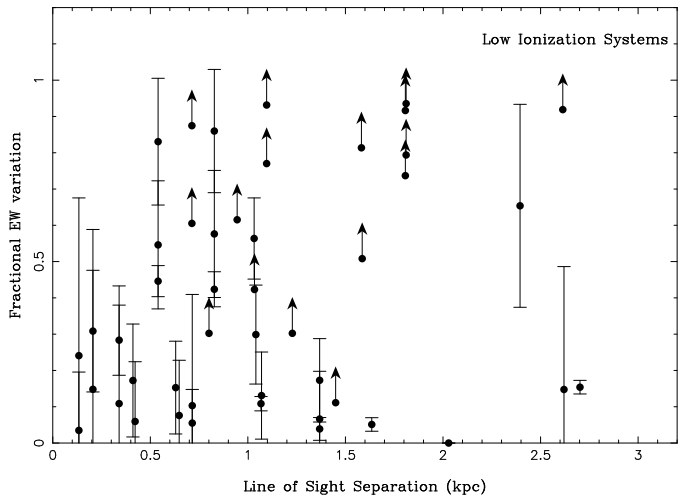}}}}
\caption{\label{frac_low}Fractional EW differences for low ionization
systems as a function of proper LOS separation}
\end{figure}

In Figures \ref{abs_low} and \ref{frac_low} we show the absolute and
fractional variations 
between sightlines as a function of LOS separation.  This is the
first time that the sizes of low ionization systems have been studied
in relatively large numbers, due to the particularly rich spectrum
and the triple nature of \apm.  
We find that the low ionization systems can have much larger 
fractional variations than the high ionization systems for LOS 
separations $> 0.5$\,\hkpc, with several cases where $\Delta EW > 80$\%.  
However, some systems exhibit little variation, even over 1\,\hkpc\ or
more.  Since Rauch et al.  (2002) found that individual low ionization
clouds can rarely be traced over more than a few hundred pc,
this is likely an effect of the combined variation of several
clouds.  Again, our inversion analysis which will be presented
in Aracil et al. will investigate this in more detail.    

Kobayashi et al (2002) have detected structural LOS differences in the Mg~II 
line of the  $z_{\rm abs}$=2.974 DLA from ground
based IR spectra taken in good seeing conditions.  This structure is
also apparent in our lower resolution spectra; the Al~II $\lambda$1670
transition is narrower in LOS B than A and C (although the Fe~II absorption
is too weak to discern such differences).  However, the absolute 
difference in EW between
the LOS is small, and consistent with those seen in other (non-damped)
low ionization systems at small separations.

\medskip

\begin{figure}
\centerline{\rotatebox{270}{\resizebox{6.5cm}{!}
{\includegraphics{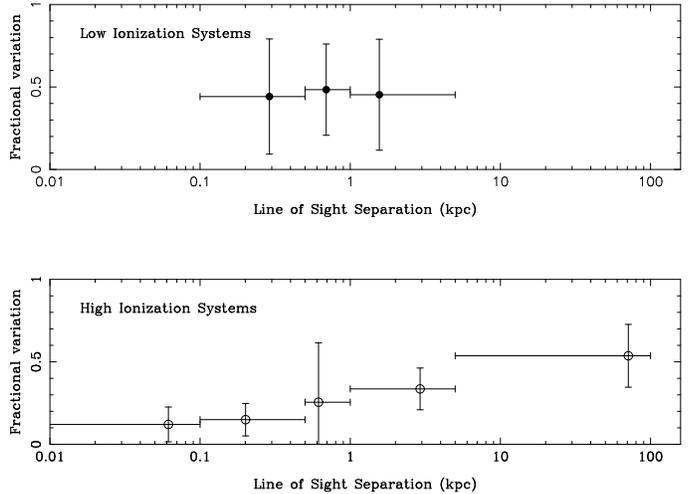}}}}
\caption{\label{frac_all}Fractional EW variations for low and high ionization
systems from this work and the literature (Lopez et al. 2000;
Rauch et al. 2001a; Rauch et al
2002; Churchill et al 2003).  We only include values based on \mgtwo, 
Fe~II, Si~IV and C~IV to be consistent with measurements towards \apm.}
\end{figure}

In Figure \ref{frac_all} we summarise the fractional
variations (based either on column density or EW, as published) in high and low
ionization systems by combining measurements made here with those
available in the literature (Lopez et al. 2000; Rauch et al. 2001a; 
Rauch et al. 2002; Churchill et al. 2003).   All the literature values 
have been converted to the cosmology adopted in the present paper.
We plot the mean fractional variation for 
each separation and the scatter (standard deviation) for that bin.  
The upper limits obtained from \apm\ (e.g. Figure \ref{frac_low}) have 
been treated as detections, with the effect that 
the scatter is slightly reduced.  However, literature upper limits
could only be included when a detection limit was quoted. It can be seen 
that for the low ionization systems the mean variation is the same for all
separations within the large scatter.  This is probably due to the fact that
individual components cannot be traced over more than a few hundred
pc.  The high ionization systems, however, show a clear trend for
larger variations with increasing LOS separation.  The scatter is
relatively small compared with the low ionization systems, except for
the 0.5 -- 1.0 kpc bin which only contains two data points.

\section{Coherence Scales of Absorption Systems}

\subsection{High Ionization Systems}

Since all of the high ionization systems detected in the
present dataset are common to all three LOS, we can only place
a lower limit on the size of the absorbers of $\sim$ 0.3 \hkpc\
(this being the largest beam separation in Table 2).
This limit is not very instructive given that observations
of wider QSO pairs have established much larger
coherence lengths for C~IV systems (e.g. Petitjean et al. 1998;
Lopez et al. 2000). More recently,
Tzanavaris \& Carswell (2003) concluded that 
that the most probable size for individual
components of C~IV complexes is $\sim 3\,h_{70}^{-1}$\,kpc,
if the `clouds' have a spherical geometry.\footnote{In their analysis,
Tzanavaris \& Carswell (2003) adopted the lensing formalism 
of Young et al. (1981) to calculate proper separations.  
However, as pointed out by Smette et al. (1992), there is an error
in the equations used by Young et al. (1981) which means that the 
LOS separations calculated by  Tzanavaris \& Carswell (2003) are 
too large by a factor of $\sim 2$. We have corrected for this
effect here.}

\subsection{Low Ionization Systems}

Since we have a modest number
of low ionization systems, several of which are only
seen in 1 or 2 of the LOS, we can employ
a maximum likelihood analysis to estimate the coherence scale
of the structures
producing them based on LOS coincidences and anti-coincidences.
This analysis relies on at least one anti-coincidence, otherwise the
likelihood function is a monotonically increasing function of cloud
radius.  Similar analyses have been performed previously for Ly$\alpha$
forest clouds and C~IV systems, but this is the first such determination for
low ionization gas halos.

We adopt the formalism of McGill (1990) with the modification
by Dinshaw et al. (1997).  One ingredient in this analysis is the
assumed geometry of the absorber, uniform spheres or flattened,
randomly inclined, disks being the two usual approximations.
There is evidence that rotation in the plane of the stellar disk of
Mg~II galaxies is reflected in the absorption gas kinematics in 
at least some cases (Steidel et al. 2002), although
absorption and emission kinematics are not
always well correlated (Lanzetta et al. 1997;
Ellison, Mallen-Ornelas \& Sawicki 2003).
More generally, the success of a simple Holmberg relation 
in describing the incidence of Mg~II absorption shows 
that a spherical gas distribution is a good approximation
(Steidel 1995), and this is the model we adopt here.
For pairs of LOS intersecting spherical clouds, the probability that
a halo is intersected by the second LOS, 
given that it is seen in the first, is given by

\begin{equation}
\label{phi}
\phi (X)=\frac{2}{\pi}\left\{ \arccos \left[ X(z)\right] - X(z)
\sqrt{1-X(z)^2} \right\}
\end{equation}

for $0 \le X(z) \le 1$ and zero otherwise.  Here, $X(z)=S(z)/2R$ where 
$S(z)$ is the LOS separation and $R$ is the absorber coherence 
radius.  
This of course assumes that all absorbers are of the same size and
that there is no column density variation within a `cloud'.
In order to make maximum use of the information available
we actually want to calculate the probability that both LOS are
intersected, if $either$ LOS shows absorption.  That is, we consider
anti-coincidences in any combination of A, B and C. 
This probability is given by

\begin{equation}
\psi=\frac{\phi}{2-\phi}
\end{equation}

\begin{figure}
\centerline{\rotatebox{270}{\resizebox{6.5cm}{!}
{\includegraphics{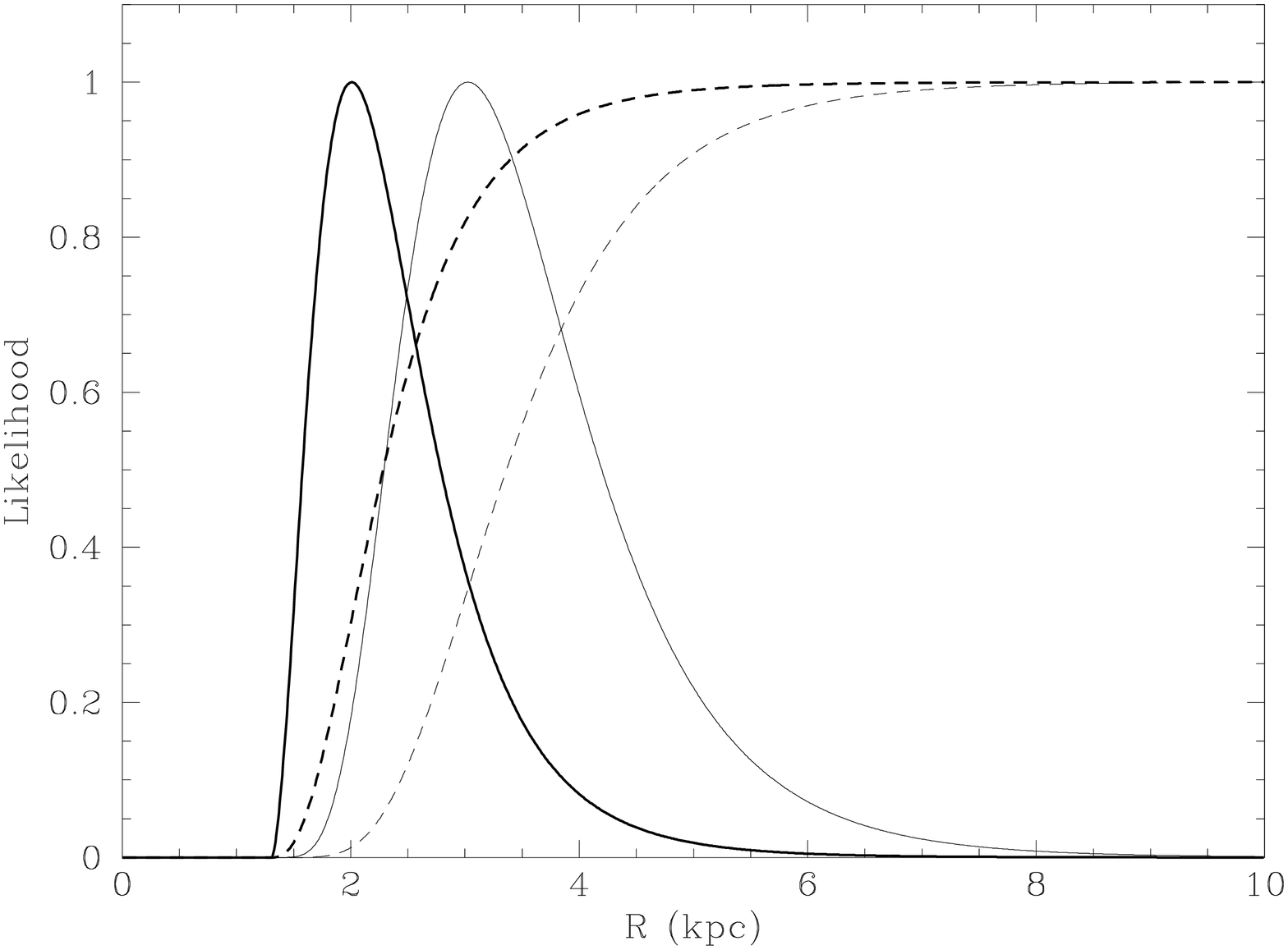}}}}
\caption{\label{max_like}
Maximum likelihood distribution (thin solid line)
and cumulative distribution (dashed line) for spherical
halos normalized to the peak value. The most likely coherence
radius is found
to be R=3.0 \hkpc\ with 95 \% confidence limits
of 2.1 and 6.2 \hkpc.  The
bold lines show the results obtained if we restrict ourselves to 
weak (EW$<$0.3 \AA) \mgtwo\ systems only, for which we deduce a
most likely scale of $R$=2.0 \hkpc\ with
95\% confidence limits  of 1.5 and 4.4 \hkpc.}
\end{figure}

And the likelihood function as a function of coherence radius
is given by

\begin{equation}
\label{like}
{\cal L}(R)=  \prod_i \psi\left[ X(z_i)\right] \prod_j \left\{1-\psi\left[
X(z_j)\right] \right\},
\end{equation}

where $i$ and $j$ denote the number of coincidences and anti-coincidences.
We note that although the original formulism of McGill (1990) was 
designed for pairs of sightlines, we can still use combinations of A, 
B and C to investigate the coherence scale of Mg~II absorption, rather
than the sizes of individual components.  
Figure \ref{max_like} shows the results of our
maximum likelihood analysis.
For the whole sample of Mg~II systems in \apm\ (16 pairs of LOS coincidences
and 13 anti-coincidences), we find
that the most probable value for the absorber coherence scale is
$R = 3.0\,h_{70}^{-1}$\,kpc, with 95\% confidence limits of 2.1 and 6.2\,kpc.  
Similarly small sizes for \mgtwo\ systems have been previously inferred from
partial coverage arguments by Petitjean et al. (2000).
This coherence scale contrasts with the much smaller sizes inferred for
individual components (Churchill et al. 2003; Rauch et al. 2002).  
However, the value $R = 3.0\,h_{70}^{-1}$\,kpc is about one order
of magnitude smaller than the dimensions of Mg~II absorbing 
halos deduced from (a) the number density of absorbers per unit redshift
and (b) the impact parameters from the QSO 
of the galaxies producing the absorption, in cases where
they have been identified (Bergeron \& Boiss\'e 1991; Steidel 1993; 
Steidel, Dickinson, \& Persson 1994).  Previous analyses of Mg~II
absorbers in lensed sightlines have also inferred scales of several
tens of kpc (Smette et al 1995).

The answer to this apparent discrepancy 
may lie in the fact that most statistical samples of Mg~II absorbers
have considered strong lines, with rest-frame equivalent widths greater
than 0.3\,\AA,
whereas many of the systems analysed here are `weak' absorbers 
(Churchill et al 1999; Rigby, Charlton \& Churchill 2002).  If we 
consider separately
the strong systems (all LOS exhibiting EW$ \geq 0.3$\,\AA), the maximum 
likelihood analysis does not converge because there are no
anti-coincidences in our sample, so that we can only infer a lower limit
to the radius/coherence scale, $R > 1.4\,h_{70}^{-1}$\,kpc
(from the largest LOS separations sampled here).  This is
also consistent with the larger sizes of strong Mg~II systems
inferred by Smette et al (1995).
For the remainder of the `weak' systems\footnote{We have,
however, excluded system 11 from the analysis due to insufficient 
S/N to detect a coincidence in LOS C at the level of the absorption
seen in LOS B.} the maximum likelihood analysis (with 8 coincidences
and 13 anti-coincidences) yields a most
probable radius of $2.0\,h_{70}^{-1}$\,kpc with 95\% confidence limits 
of 1.5 and 4.4\,kpc (see Figure \ref{max_like}).   If the adopted
lens redshift is in fact $z_{lens} < 1.062$
this would imply even smaller beam separations
(e.g. by $\sim$50\% for $z_{lens}=0.6$), so our conclusion that
weak absorbers have small sizes would remain unchanged.

For the DLA at $z_{\rm abs} = 2.974$, where all absorption lines
are seen in all three sight-lines, we similarly deduce a lower
limit to the size of 0.34\,\hkpc. This is in agreement with the size
deduced by Petitjean et al. (2000) based on the absence of 
partial coverage in the \lya\ absorption line.

\section{Velocity Shear}

\begin{figure}
\centerline{\rotatebox{270}{\resizebox{6.5cm}{!}
{\includegraphics{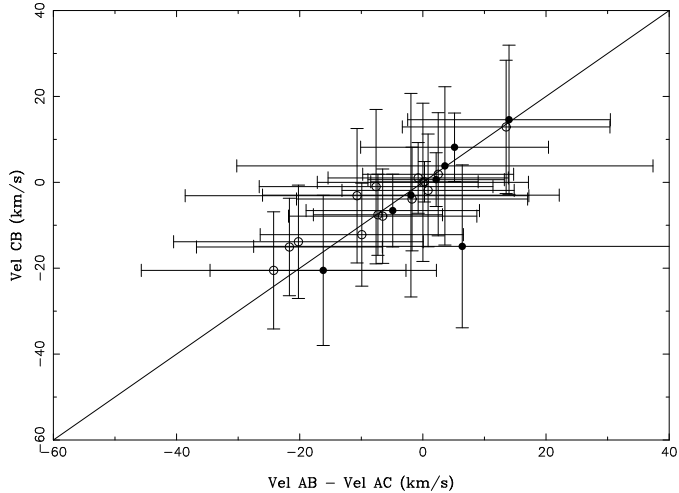}}}}
\caption{\label{shear_check}Consistency check for cross-correlation shear
velocities.  We compare the difference between the cross correlation
velocities for $v_A - v_B$ and $v_B - v_C$ with that computed for $v_B - v_C$.}
\end{figure}

Velocity shear between sightlines was calculated by cross-correlating
the absorption line profiles for each LOS combination using 
the task {\tt fxcor} in IRAF.  
Errors within this package are calculated using the
formalism of Tonry \& Davis (1979).  An independent error check
can be carried out using the fact that shifts between each LOS combination
form a closed system.  That is, the difference in the calculated
shifts between A--B and A--C should equal the shift between C--B.
We plot the results of this consistency check in Figure \ref{shear_check}.
It can be seen that in most cases the agreement is very good,
and well within the formal error bars.

\subsection{High Ionization Systems}

\begin{figure}
\centerline{\rotatebox{270}{\resizebox{6.5cm}{!}
{\includegraphics{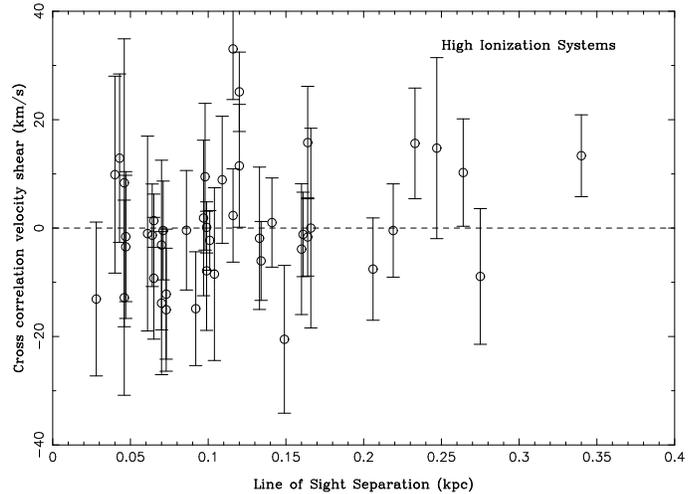}}}}
\caption{\label{shear_high}Velocity shear as a function of separation for
high ionization systems.}
\end{figure}

In Figure \ref{shear_high} we present the shear velocities as a function of
LOS separation for high ionization systems.  We find that the
shear is generally less than 20\,\kms\ for separations 
up to $\sim$0.35 \hkpc.
Again, this complements the results by Rauch et al (2001a) who had
few points below 0.2 kpc.  With these additional points
at small separations, 
it can now be seen that shear velocities in high
ionization systems are generally less than 15--20 \kms\ at separations 
smaller than 0.5 \hkpc, with evidence for an increase in shear at larger
separations (Rauch et al. 2001a).

\subsection{Low Ionization Systems}

\begin{figure}
\centerline{\rotatebox{270}{\resizebox{6.5cm}{!}
{\includegraphics{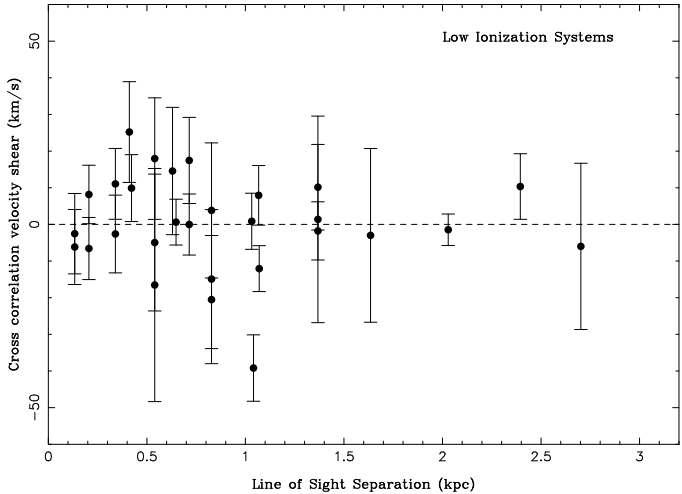}}}}
\caption{\label{shear_low}Velocity shear as a function of separation for
low ionization systems.}
\end{figure}

The velocity shear for low ionization systems is shown in Figure 
\ref{shear_low}.  
We do not find any evidence for the significant
increase in shear on scales greater than 1\,kpc seen in C~IV systems 
by Rauch et al. (2001a).  
The observed shear between the metal lines of the DLA is also small, 
less than 10\,\kms, consistent with the those of the other, weaker, 
low ionization absorbers.
However, our analysis is limited by the fact that
we have insufficient spectral resolution to perform a column density
weighted calculation (c.f. Rauch et al. 2002; Churchill et al. 2003).
This is particularly important for the
low ionization
systems where the scale of individual components
is much smaller than our proper beam separation.  This is
highlighted by the \mgtwo\ system at $z=1.813$ where 
the absorption lines are of
of similar strengths in LOS A and B, 
but are much weaker in the spatially intermediate LOS C.
Our cross-correlation velocity shear therefore traces the shift between
the strongest components in each LOS.

\section{Chemical Uniformity}

In cases where two ions are observed in more than one
LOS (and the lines have EW$<$0.2 \AA\ so as to permit a column density
determination), we can test for transverse variations 
in the ratio of their column densities.  
Such variations may result from differences between the
sightlines in chemical composition, dust depletion
and ionization conditions, or any combination of these.
However, since we typically only observe two elements
in each absorption system,
we do not have sufficient information
to disentangle these effects.

There are two high ionization systems for which we can study
transverse abundance differences through the C~IV/Si~IV ratio.  
Table \ref{civ2siiv} summarises 
these results; relative column densities for weak high ionization 
systems agree to within 0.1\,dex over a few hundred pc.  
Since we have found that
high ionization systems vary little between LOS, it is not
surprising that their column density ratios are in good agreement too.
Unfortunately, we can not investigate differences in the low
ionization systems (apart from the DLA)
since all the \mgtwo\ systems which have associated
Fe~II are too strong to determine accurate column densities.

For the DLA (or possible sub-DLA), we have available low ionization 
lines of Al~II and Fe~II.
Although Al~II $\lambda$1670 is usually saturated in DLAs, we see
no evidence for this in the HIRES spectrum and the lack of flat-bottomed
profiles argues against de-saturation by partial coverage.  
This is not unexpected given the relatively low $N$(HI) of this
DLA (Petitjean et al 2000).  Since Al~II and Fe~II represent the major 
ionisation stages of these elements in DLAs (and ionisation corrections are
relatively small even for sub-DLAs, see Dessauges-Zavadsky et al. 2003),
we can use the EWs in Table \ref{ew_table} to deduce abundance ratios
[Al/Fe]\,$=  +0.15^{+0.09}_{-0.12}$, $-0.04^{+0.09}_{-0.13}$, $+0.05^{+0.15}_{-0.22}$ in A, B and C respectively\footnote{Using the conventional notation 
[X/Y] = log [$N$(X)/$N$(Y)]$_{\rm DLA}$ $-$ log [$N$(X)/$N$(Y)]$_{\odot}$.}.  
Although these values may suggest 
mild variations of the [Al/Fe] ratio between the three sight-lines,
they are actually all consistent with the solar
value within the errors \footnote{Adopted solar abundances
are meteoritic values taken from Grevesse \& Sauval (1998)}.  
Chemical uniformity in absorbers
has been previously pointed out both along (Prochaska 2003) and
across (Churchill et al. 2003) quasar lines of sight.  Our observations
concur with these findings, with little detected variation over
$\sim$ 350 pc.  

\section{Discussion}

In the Milky Way, the warm, highly ionized gas component is relatively 
smoothly distributed with little or no change in column density over 
several degrees on the sky; SiIV and CIV absorptions in particular 
are generally well matched (Savage, Sembach \& Lu 1997).
In contrast, very small-scale variations (on the order of a few AUs)
have been observed in the cold ($\sim$ 100 K) neutral gas component 
(traced by Na~I) of the interstellar medium (e.g. Andrews, Meyer 
\& Lauroesch 2001; Lauroesch, Meyer \& Blades 2000). 
The singly ionized species
of Fe~II and Mg~II studied here are intermediate between these two phases
and most likely occur in a warm neutral medium.  These species
trace each other well in terms of velocity structure, as well as showing
a close kinematic relation to Ca~II which is more often studied in 
the local ISM 
(e.g. Welty, Morton \& Hobbs 1996; Redfield \& Linsky 2002).
Such low ionization species do exhibit some variation on scales
of a few tens to tens of thousands of AU, but the LOS differences are
less marked than in the cold phase traced by Na~I  (e.g. Meyer \& Roth 1991; 
Meyer \& Blades 1996; Lauroesch et al 1998; Price, Crawford \& 
Howarth 2001).  Variation in Ca~II and other low ions can be seen both 
in terms of velocity structure and column density, although
such variation is by no means ubiquitous (e.g. Rauch et al. 
2002).   The presence of these small scale variations can be 
explained by density
contrast within a diffuse cloud where the bulk of the low ionization
gas is concentrated in the highest density peaks (Lauroesch \& 
Meyer 2003).  The variation of different species between lines of sight 
(or temporally in a single Galactic sightline) is therefore dependent 
not only on the size of the `cloud' itself, but also on the local physical 
conditions that maintain the ionization balance.

Qualitatively similar trends have been observed in high redshift absorbers.
On scales of a few pc the high ionization gas is effectively featureless 
(Rauch et al. 1999), with variations typically  $<$30\% for transverse
separations of 50 pc -- 5 \hkpc\ (Rauch et al. 2001a; this work).
However, the location of this high ionization gas remains uncertain.
The ubiquity of C~IV absorption associated with the \lya\ forest,
even at low H~I column densities (e.g. Ellison et al. 2001), and 
coherence lengths of more than 100 \hkpc\ (Petitjean et al 1998;
Lopez, Hagen \& Reimers 2000) suggest an intergalactic origin.
However, Adelberger et al. (2003) find a strong correlation between Lyman
break galaxies and C~IV absorption out to impact parameters of 
hundreds of kpc, indicating a link between galaxy winds and metal 
enriched gas.  A similar conclusion was reached by Rauch et al.
(2001a) based on the small
internal turbulence of C~IV systems.

Similar to the situation  in Galactic sightlines, high redshift
absorbers exhibit much stronger variation in the low ionization species.
Given the pervasive small-scale structure observed in the Milky Way,
it is not surprising that individual \mgtwo\ components
cannot be traced between sight-lines separated by a few tens to a few
hundred pc (Rauch et al. 1999, 2002) and that large fractional
EW differences are common in these systems (this work).  
However, coincidences of absorption between LOS indicate that there
is coherence in the absorption structure on somewhat larger scales.
This can be compared with the `clumpy but coherent' structure seen
in Mg~II and other low ions in the local ISM out to at least 100 pc
(Redfield \& Linksy 2002).  Based on
a maximum likelihood analysis, we have been able to infer the likely
coherence length of weak \mgtwo\ systems (as opposed to the size of
individual components)
$R \sim 2$ \hkpc.  This coherence scale must be understood 
in context  with the overall galaxy 
sizes inferred for strong \mgtwo\ systems (e.g. Steidel 1995) which
show absorption out to several tens of kpc from the absorbing galaxies.

One possibility is that
the weak absorbers represent a distinct population whose actual sizes 
are a few kpc, comparable to those of present-day dwarf galaxies.  
Compared with strong \mgtwo\ absorbers, the weak systems in our sample
have relatively simple kinematic structure and 
are spread over narrow velocity ranges. 
In generic cold dark matter models of galaxy formation
(e.g. Mo \& Miralda-Escude 1996) 
there is a steep correlation between circular velocity and
halo dimension which agrees with the small sizes
derived from our maximum likelihood analysis.  However, the small
sizes inferred here contradict the suggestion by Churchill \& Le Brun 
(1998) that weak \mgtwo\ absorbers are `giant' LSB systems, although
low luminosities remain highly feasible.

Alternatively, the
weak \mgtwo\ systems may occur in sightlines through large galaxies,
but in regions where interstellar clouds have lower filling factors
leading to smaller inferred coherence lengths.
Indeed, Rigby et al. (2002) suggest that although single cloud
weak \mgtwo\ systems (see below) may arise from a distinct
population, the multi-component systems are likely have the
same origin as the strong clouds.
In this picture, the absorbing clouds are made up of pc-scale components 
with a total covering factor of about unity (Srianand \& Khare 1994; 
Rauch et al 2002) in the central few tens of kpc of the galaxy.   
Further out, the filling factor is smaller, so that anti-coincidences
may be observed indicating smaller
sizes.  Inferred scales of $2R \sim$ 4 \hkpc\ compared to $\sim$ 40 kpc 
(Steidel 1995) would imply filling factors that are an order of magnitude
smaller for the weak \mgtwo\ systems.  Smaller filling factors would 
also explain the tendency towards fewer components in low EW systems.
The above scenario is qualitatively born out by an observed 
anti-correlation between \mgtwo\ column density and galaxy impact 
parameter (Churchill et al. 2000).  Indeed, Churchill et al. (2000) find
that all high impact parameter intersections have small EW \mgtwo\ 
absorbers.  This scenario could be tested with
an imaging and spectroscopic campaign to identify the
galaxies responsible for absorption.

Finally, we comment on the observation that 
two thirds of weak Mg~II absorbers
have been previously found to be single cloud systems (Rigby et al. 2002). 
The HIRES spectra of APM08279+5255 (Ellison et al. 1999b)
show this not to be the case here;  
all of the eight weak Mg~II systems we have identified 
exhibit multiple velocity components when observed with HIRES.
Bearing in mind that the HIRES observations do not resolve
the three images, this multicomponent structure is probably
a reflection of the small scale variations seen by Rauch et al. (2002).
Moreover, the exceptional S/N of the HIRES spectrum of APM08279+5255
reveals components that would not normally
be identified in most spectra.  The Mg~II system at $z_{\rm abs}$=1.291,
which has an unresolved main component and a weaker, broader feature
$\sim$ 25 \kms\ away (Ellison et al. 1999b),
would have been classified by Rigby et al. (2002) as a single cloud
weak absorber in a lower S/N spectrum.  Using ionization models, Rigby
et al. found the sizes of single cloud
absorbers \textit{with Fe~II} to be $\sim10$ pc.  The models for
weak \mgtwo\ systems without Fe~II are less well constrained, allowing
for cloud sizes which vary between a few pc and several tens of kpc.  
Since the system at $z_{\rm abs}$=1.291 would have
been classified as an (albeit, Fe~II free) weak single cloud
absorber in spectra of S/N typical of the data used by Rigby et al., 
we can now significantly improve upon the size estimate of this type 
of absorber with the
STIS data.  Given that absorption is detected in both LOS A and B 
(in C the S/N is insufficient to provide useful information)
we deduce a minimum size of 2.4 $h_{70}^{-1}$ kpc, more
than 20 times larger than the value estimated 
(indirectly) for the Fe-rich systems.

\section{Conclusions}

In this paper, 
we have presented moderate resolution STIS spectra 
of the tripled imaged QSO APM08279+5255,
and focussed our analysis on an 
investigation of the sizes and coarse structure
of intervening metal absorption systems.  
By capitalising upon the unusually absorption-rich lines of sight 
offered by the triply imaged QSO \apm, we have significantly 
increased the number of
high ionization systems (primarily  C~IV) probed at small ($<0.1$\,\hkpc)
separations.  This study is also the most comprehensive to date
of the variation of low ionization (e.g. Mg~II) systems.  Our
principal conclusions are as follows.

\begin{enumerate}

\item  High ionization systems generally show small EW variations
on scales of a few tens to a few hundreds of pc, although we
find evidence in at least one system for small component differences
across the LOS.  This complements
the work of Rauch et al. (2001a) whose observations probed mostly larger
separations for C~IV systems in three pairs of lensed QSOs.  
Taken together, all of these data 
indicate a steady increase in
fractional variation with increasing LOS separation, although
rarely exceeding 40\%.  All the high ionization systems
occur in all three lines-of-sight to \apm,
implying sizes of $> 3$\,\hkpc.

\item  In contrast to the high ionization systems, low ionization 
complexes exhibit significant variation ($\Delta$~EW $> 80$\%) on scales 
greater than a few hundred pc, although we lack the spectral resolution to 
trace individual clouds between sightlines.   There is no trend of
fractional variation as a function of LOS separation, which can be
explained by the small sizes ($<$ few hundred pc) of individual components.
For strong (EW$>$ 0.3 \AA)
Mg~II systems we can only infer a minimum radius of $\sim$ 3 \hkpc,
although the actual size is likely to be significantly larger
(Steidel 1995; Smette et al 1995). For weaker systems (EW$<$ 0.3 \AA), 
we apply a 
maximum likelihood method and infer a most 
probable coherence scale of 2.0\,$h_{70}^{-1}$ kpc with 95\% 
confidence limits 
of 1.5 and 4.4\,kpc.  Therefore, we suggest that these are either a
population distinct from strong \mgtwo\ systems,
perhaps associated with dwarf galaxies,
or that they occur
in the outer regions of large, luminous galaxies where 
their filling factor is lower.

\item  The velocity shear for both high and low ionization systems
is small, typically less than 20 \kms.  However, since we can not
trace individual low ionization components, the shear velocities
for these systems reflects mainly the shift in the location 
of the strongest component(s).

\item  For two of the high ionization systems where the lines
are opticall thin and the EWs can be converted to column
densities, we find
consistent ratios of $N$(C~IV)/$N$(Si~IV) between LOS.  The abundance
ratio of [Al/Fe] in the DLA (or possible sub-DLA) also exhibits only 
mild variation ($< 0.1$\,dex) and is approximately solar in the three LOS.

\item  We note a deficit of single cloud weak \mgtwo\ absorbers
in the HIRES spectrum of \apm\ 
and speculate that this may be due to the superposition
of sightlines in the spatially unresolved ground-based data.  
This and other issues will
be addressed in a future paper (Aracil et al, in preparation) that
uses an inversion analysis to produce high spectral resolution
for each LOS by combining information from the STIS and HIRES 
spectra.

\end{enumerate}

\begin{acknowledgements}

We are grateful to our $HST$ program coordinator, Beth Perriello at STScI,
for continued support through the observations, and to Steven Smartt
at the UK HST user support facility which was based at the 
Institute of Astronomy in Cambridge.  This work benefitted from
comments and suggestions from Bob Carswell and Alain Smette.

\end{acknowledgements}

%
%

\begin{table*}
\begin{center}
\begin{tabular}{lcccccc} \hline \hline
$\lambda_0$  &  Range(\AA) & Date & Total Exp Time(s)  & \multicolumn{3}{c}{S/N} \\ 
 & & & & LOS A & LOS B & LOS C \\ \hline
6252 & 5970--6530  & Nov 10 2001  &  14\,900 & 60 & 45 & 25 \\
6768 & 6490--7050  & Dec 26 2001  &  11\,800 & 60 & 45 & 25 \\
7283 & 7000--7560  & Dec 26-27 2001  &  11\,800 & 60 & 40 & 25\\
7795 & 7500--8070  & Dec 27 2001, Nov 25 2002  &  14\,900 & 40 & 35 & 20\\
8311 & 8030--8600  & Nov 25-26 2002  &  21\,100 & 30 & 25 & 17 \\ \hline
\end{tabular}
\caption{\label{obs_table} Log of STIS Observations}
\end{center}
\end{table*}

\begin{table*}
\begin{center}
\begin{tabular}{lcccc} \hline \hline
Absorber No. &Transition  & Redshift  & A-B linear & A-C linear \\ 
& & & sep$^{n}$ (kpc h$_{70}^{-1}$) & sep$^{n}$ (kpc h$_{70}^{-1}$)\\ \hline
1  & Mg~II $\lambda\lambda$2796, 2803  &  1.181  & 2.702  &  1.067 \\ 
2  & Si~IV $\lambda\lambda$1393, 1402  &  3.379  & 0.164  &  0.065 \\
3  & C~IV  $\lambda\lambda$1548, 1550  &  2.974  & 0.340  &  0.134 \\
4  & Mg~II $\lambda\lambda$2796, 2803  &  1.209  & 2.620  &  1.034 \\
5  & Mg~II $\lambda\lambda$2796, 2803  &  1.211  & 2.614  &  1.032 \\
6  & Si~IV $\lambda\lambda$1393, 1402  &  3.502  & 0.120  &  0.047 \\
7  & Si~IV $\lambda\lambda$1393, 1402  &  3.514  & 0.116  &  0.046 \\
8  & C~IV  $\lambda\lambda$1548, 1550  &  3.109  & 0.275  &  0.109 \\
9  & Fe~II $\lambda$1608               &  2.974  & 0.340  &  0.134 \\
10 & C~IV  $\lambda\lambda$1548, 1550  &  3.134  & 0.264  &  0.104 \\
11 & Mg~II $\lambda\lambda$2796, 2803  &  1.291  & 2.395  &  0.945 \\
12 & C~IV  $\lambda\lambda$1548, 1550  &  3.171  & 0.247  &  0.098 \\
13 & C~IV  $\lambda\lambda$1548, 1550  &  3.203  & 0.233  &  0.092 \\
14 & C~IV  $\lambda\lambda$1548, 1550  &  3.239  & 0.219  &  0.086 \\
15 & Fe~II $\lambda$1608               &  1.550  & 1.811  &  0.715 \\
16 & Fe~II $\lambda$1608               &  1.552  & 1.807  &  0.713 \\
17 & Al~II $\lambda$1670               &  2.974  & 0.340  &  0.134 \\
18 & Fe~II $\lambda$1608               &  1.813  & 1.368  &  0.540 \\
19 & C~IV  $\lambda\lambda$1548, 1550  &  3.379  & 0.164  &  0.065 \\
20 & C~IV  $\lambda\lambda$1548, 1550  &  3.386  & 0.161  &  0.064 \\
21 & Si~IV $\lambda\lambda$1393, 1402  &  3.893  & 0.005  &  0.002 \\
22 & Mg~II $\lambda\lambda$2796, 2803  &  1.444  & 2.029  &  0.801 \\
23 & Si~IV $\lambda\lambda$1393, 1402  &  3.913  & ...    &  ...   \\
24 & Si~IV $\lambda\lambda$1393, 1402  &  3.917  & ...    &  ...   \\
25 & C~IV  $\lambda\lambda$1548, 1550  &  3.502  & 0.120  &  0.047 \\
26 & C~IV  $\lambda\lambda$1548, 1550  &  3.514  & 0.116  &  0.046 \\
27 & C~IV  $\lambda\lambda$1548, 1550  &  3.558  & 0.101  &  0.040 \\
28 & Mg~II $\lambda\lambda$2796, 2803  &  1.550  & 1.811  &  0.715 \\
29 & Mg~II $\lambda\lambda$2796, 2803  &  1.552  & 1.807  &  0.713 \\
30 & C~IV  $\lambda\lambda$1548, 1550  &  3.655  & 0.071  &  0.028 \\
31 & Fe~II $\lambda$2600               &  1.813  & 1.368  &  0.540 \\
32 & C~IV  $\lambda\lambda$1548, 1550  &  3.857  & 0.014  &  0.005 \\
33 & C~IV  $\lambda\lambda$1548, 1550  &  3.893  & 0.005  &  0.002 \\
34 & C~IV  $\lambda\lambda$1548, 1550  &  3.899  & 0.003  &  0.001 \\
35 & C~IV  $\lambda\lambda$1548, 1550  &  3.912  & ...    &  ...   \\
36 & C~IV  $\lambda\lambda$1548, 1550  &  3.917  & ...    &  ...   \\
37 & Mg~II $\lambda\lambda$2796, 2803  &  1.813  & 1.368  &  0.540 \\
38 & Ca~II $\lambda\lambda$3934, 3969  &  1.062  & 3.086  &  1.218 \\
39 & Mg~II $\lambda\lambda$2796, 2803  &  2.042  & 1.070  &  0.422 \\
40 & Mg~II $\lambda\lambda$2796, 2803  &  2.067  & 1.041  &  0.411 \\ \hline
\end{tabular}
\caption{\label{system_table} A list of intervening Mg~II, C~IV and Si~IV absorbers.
Proper linear separations in are calculated assuming  the lens  at  $z$=1.06,
A--C separation = 0.15 arcsec, A--B separation = 0.38 arcsec
with
$\Omega_M$=0.3, $\Omega_{\Lambda}$=0.7.  
The transition numbers refer to tick marks in Figure \ref{stis_spec}. }
\end{center}
\end{table*}

\begin{landscape}
\begin{table}
\begin{center}
\begin{tabular}{lcccccccccc} \hline \hline
No. & Absorber  & Redshift & \multicolumn{2}{c}{LOS A} & \multicolumn{2}{c}{LOS B} & \multicolumn{2}{c}{LOS C}& \multicolumn{2}{c}{HIRES} \\ 
& &  & EW$_1$ &  EW$_2$ & EW$_1$ &  EW$_2$ & EW$_1$ &  EW$_2$ & EW$_1$ &  EW$_2$\\ \hline

1 & MgII  & 1.181 & 2.57$\pm$0.04 & 2.26$\pm$0.02 &  3.03$\pm$0.04 & 2.98$\pm$0.02 &  2.88$\pm$0.04 & 2.68$\pm$0.02 &  2.595 & 2.433 \\  
2 & SiIV  & 3.379 & ...           & 0.08$\pm$0.02 &  ...           & 0.09$\pm$0.02 &  ...           & 0.09$\pm$0.02 &  ...   & 0.095 \\    
3 & CIV   & 2.974 & 0.17$\pm$0.01 & 0.10$\pm$0.01 &  0.17$\pm$0.02 & 0.08$\pm$0.01 &  0.17$\pm$0.03 & 0.06$\pm$0.02 &  0.167 & 0.076 \\   
4 & MgII  & 1.209 & 0.05$\pm$0.01 & 0.03$\pm$0.01 &  0.06$\pm$0.02 & 0.04$\pm$0.02 &  $<$0.03       & $<$0.03       &  0.051 & 0.034 \\  
5 & MgII  & 1.211 & 0.37$\pm$0.01 & 0.34$\pm$0.01 &  $<$0.03       & $<$0.03       &  0.16$\pm$0.04 & 0.25$\pm$0.04 &  0.234 & 0.236 \\  
6 & SiIV  & 3.502 & 0.14$\pm$0.01 & 0.09$\pm$0.01 &  0.18$\pm$0.01 & 0.09$\pm$0.01 &  0.14$\pm$0.02 & 0.08$\pm$0.02 &  0.145 & 0.087 \\  
7 & SiIV  & 3.514 & 0.04$\pm$0.01 & 0.03$\pm$0.01 &  0.05$\pm$0.01 & 0.02$\pm$0.01 &  0.06$\pm$0.02 & $<$0.03       &  0.043 & 0.018 \\  
8 & CIV   & 3.109 & 0.31$\pm$0.01 & 0.18$\pm$0.01 &  0.28$\pm$0.01 & 0.15$\pm$0.01 &  0.30$\pm$0.02 & 0.13$\pm$0.02 &  0.308 & 0.175 \\  
9 & FeII  & 2.974 & 0.04$\pm$0.01 & ...           &  0.05$\pm$0.01 & ...           &  0.05$\pm$0.02 & ...           &  0.041 & ...   \\  
10 & CIV  & 3.134 & 0.14$\pm$0.01 & 0.08$\pm$0.01 &  0.16$\pm$0.02 & 0.12$\pm$0.01 &  0.17$\pm$0.03 & 0.11$\pm$0.03 &  0.143 & 0.080 \\  
11 & MgII & 1.291 & 0.08$\pm$0.01 & 0.05$\pm$0.01 &  0.03$\pm$0.01 & $<$0.03       &  $<$0.03       & $<$0.03       &  0.046 & 0.028 \\  
12 & CIV  & 3.171 & 0.26$\pm$0.01 & 0.13$\pm$0.01 &  0.22$\pm$0.02 & 0.08$\pm$0.02 &  0.32$\pm$0.03 & 0.15$\pm$0.03 &  0.242 & 0.131 \\  
13 & CIV  & 3.204 & ...           & 0.09$\pm$0.01 &  ...           & 0.10$\pm$0.02 &  ...           & 0.10$\pm$0.03 &  ...   & 0.098 \\  
14 & CIV  & 3.239 & 0.04$\pm$0.01 & 0.03$\pm$0.01 &  0.05$\pm$0.01 & 0.02$\pm$0.01 &  0.07$\pm$0.02 & $<$0.03       &  0.051 & 0.027 \\  
15 & FeII & 1.550 & 0.10$\pm$0.02 & ...           &  $<$0.02       & ...           &  0.09$\pm$0.03 & ...           &  ...   & ...   \\  
16 & FeII & 1.552 & 0.08$\pm$0.01 & ...           &  $<$0.02       & ...           &  $<$0.03       & ...           &  0.041 & ...   \\  
17 & AlII & 2.974 & 0.19$\pm$0.01 & ...           &  0.14$\pm$0.01 & ...           &  0.20$\pm$0.03 & ...           &  0.202 & ...   \\  
18 & FeII & 1.813 & 0.18$\pm$0.01 & ...           &  0.21$\pm$0.02 & ...           &  0.03$\pm$0.03 & ...           &  ...   & ...   \\  
19 & CIV  & 3.379 & 0.37$\pm$0.02 & ...           &  0.33$\pm$0.02 & ...           &  0.32$\pm$0.03 & ...           &  0.352 & ...   \\  
20 & CIV  & 3.386 & ...           & 0.11$\pm$0.02 &  ...           & 0.12$\pm$0.02 &  ...           & 0.10$\pm$0.02 &  ...   & 0.120 \\  
21 & SiIV & 3.893 & 0.35$\pm$0.01 & 0.28$\pm$0.01 &  0.36$\pm$0.01 & 0.28$\pm$0.01 &  0.34$\pm$0.02 & 0.28$\pm$0.02 &  0.364 & 0.284 \\  
22 & MgII & 1.444 & 0.04$\pm$0.01 & ...           &  0.04$\pm$0.01 & ...           &  $<$0.03       & ...           &  0.045 & ...   \\  
23 & SiIV & 3.913 & 0.33$\pm$0.01 & 0.20$\pm$0.01 &  0.32$\pm$0.01 & 0.21$\pm$0.01 &  0.26$\pm$0.02 & 0.27$\pm$0.02 &  ...   & ...   \\  
24 & SiIV & 3.917 & ...           & 0.02$\pm$0.01 &  ...           & 0.02$\pm$0.01 &  ...           & 0.04$\pm$0.02 &  ...   & ...   \\  
25 & CIV  & 3.502 & 0.10$\pm$0.01 & ...           &  0.13$\pm$0.01 & ...           &  0.13$\pm$0.03 & ...           &  0.113 & ...   \\  
26 & CIV  & 3.514 & 0.03$\pm$0.01 & 0.01$\pm$0.01 &  0.05$\pm$0.01 & 0.03$\pm$0.01 &  0.05$\pm$0.02 & $<$0.03       &  0.043 & 0.028 \\  
27 & CIV  & 3.558 & 0.05$\pm$0.01 & 0.02$\pm$0.01 &  0.04$\pm$0.01 & 0.03$\pm$0.01 &  0.04$\pm$0.02 & $<$0.03       &  0.040 & 0.026 \\  



28 & MgII & 1.550 & 0.31$\pm$0.01 & 0.24$\pm$0.01 &  $<$0.02       & $<$0.02       &  0.29$\pm$0.03 & 0.28$\pm$0.03 &  0.229 & 0.151 \\  
29 & MgII & 1.552 & 0.24$\pm$0.01 & 0.20$\pm$0.01 &  $<$0.02       & $<$0.02       &  $<$0.03       & $<$0.03       &  0.119 & 0.102 \\  
30 & CIV  & 3.655 & 0.06$\pm$0.01 & 0.03$\pm$0.01 &  0.07$\pm$0.02 & 0.03$\pm$0.02 &  0.06$\pm$0.01 & 0.03$\pm$0.01 &  0.061 & 0.034 \\  
31 & FeII & 1.813 & 0.14$\pm$0.01 & ...           &  0.15$\pm$0.02 & ...           &  0.06$\pm$0.02 & ...           &  0.107 & ...   \\  
32 & CIV  & 3.857 & 0.20$\pm$0.01 & 0.11$\pm$0.01 &  0.19$\pm$0.02 & 0.09$\pm$0.02 &  0.23$\pm$0.02 & 0.11$\pm$0.02 &  0.217 & 0.106 \\  
33 & CIV  & 3.893 & 0.66$\pm$0.05 & ...           &  0.64$\pm$0.05 & ...           &  0.65$\pm$0.05 & ...           &  0.664 & ...   \\  
34 & CIV  & 3.899 & ...           & 0.70$\pm$0.04 &  ...           & 0.66$\pm$0.04 &  ...           & 0.70$\pm$0.05 &  ...   & ...   \\  
35 & CIV  & 3.912 & 0.63$\pm$0.03 & 0.40$\pm$0.03 &  0.59$\pm$0.03 & 0.36$\pm$0.03 &  0.65$\pm$0.05 & 0.39$\pm$0.04 &  ...   & ...   \\  
36 & CIV  & 3.917 & 0.49$\pm$0.03 & 0.41$\pm$0.03 &  0.43$\pm$0.03 & 0.34$\pm$0.03 &  0.50$\pm$0.03 & 0.36$\pm$0.03 &  ...   & ...   \\  
37 & MgII & 1.813 & 0.80$\pm$0.02 & 0.58$\pm$0.02 &  0.77$\pm$0.02 & 0.48$\pm$0.02 &  0.44$\pm$0.03 & 0.21$\pm$0.03 &  0.686 & 0.477 \\  
38 & CaII & 1.062 & ...           & ...           &  ...           & ...           &  ...           & ...           &  ...   & ...   \\  
39 & MgII & 2.041 & 0.21$\pm$0.02 & 0.14$\pm$0.02 &  0.24$\pm$0.02 & 0.13$\pm$0.02 &  0.22$\pm$0.03 & 0.18$\pm$0.03 &  0.222 & 0.154 \\  
40 & MgII & 2.066 & 0.31$\pm$0.04 & ...           &  0.45$\pm$0.04 & ...           &  0.38$\pm$0.04 & ...           &  0.297 & ...   \\  \hline

\end{tabular}
\caption{\label{ew_table} Rest frame EW measurements (values for
doublet components indicated by subscripts 1 and 2) for STIS and HIRES 
spectra. Upper limits are 3 $\sigma$.  No data (`...') is given if
the line is known from the HIRES spectrum to be contaminated or blended,
or the line is not a multiplet. }
\end{center}
\end{table}
\end{landscape}

\begin{table}
\begin{tabular}{lccc} \hline \hline
Redshift & LOS A ratio &  LOS B ratio &  LOS C ratio \\ \hline
3.502 & 0.22$\pm$0.06 & 0.18$\pm$0.06 & 0.29$\pm$0.14 \\
3.514 & 0.27$\pm$0.13 & 0.33$\pm$0.16 & 0.21$\pm$0.17 \\ \hline
\end{tabular}
\caption{\label{civ2siiv}Log N(C~IV)/log N(Si~IV) column density ratios 
for systems with EW$<$0.2 \AA\ for each line of sight.}
\end{table}

\end{document}